\documentclass[12pt,a4paper]{article}
\usepackage{fullpage,epsfig,graphics,amsbsy,amssymb,cancel,slashed,mathrsfs,tensor}
\usepackage{dsfont}
\usepackage{float}
\usepackage{amsfonts}
\usepackage{psfrag,hyperref}
\usepackage{graphicx,color}
\usepackage{wrapfig}
\usepackage{physics}

\usepackage{tikz-cd,tikz}
\usetikzlibrary{calc}
\usetikzlibrary{decorations.pathreplacing}

\newcommand{\be}{\begin{eqnarray}}
\newcommand{\ee}{\end{eqnarray}}
\usepackage{amsmath}
\usepackage{slashed}
\numberwithin{equation}{section}
\usepackage{caption}
\usepackage{subcaption} 
\usepackage[nosort]{cite}

\graphicspath{{./figures/}}

\def\N{{\cal N}}

\def\E{{\cal E}}

\newcommand{\bea}{\begin{eqnarray}}
\newcommand{\eea}{\end{eqnarray}}  
\newcommand{\nn}{\nonumber}
\renewcommand{\Tr}{\textrm{Tr}}

\newcommand{\NN}{\mathcal{N}}

\newcommand{\Li}{\mbox{Li}}

\newcommand{\OO}{{\mathcal O}}
\newcommand{\ZZ}{{\mathcal Z}}

 \newcommand{\SU}{\mathrm{SU}}

\newcommand{\dsl}{\slashed{\partial}}

\def\al{\alpha}
\def\om{\omega}

\newcommand{\bP}{\mathbf{P}}
\newcommand{\bQ}{\mathbf{Q}}

\usepackage{bbold}

\graphicspath{{./figures/}}

\newcommand{\io}{\textrm{i}}
\newcommand{\wt}[1]{\widetilde{#1}}
\newcommand{\bkt}[1]{\left( #1 \right)}

\begin{document}

\thispagestyle{empty}
\begin{flushright} \small
UUITP-34/25\\
MIT-CTP/5968
 \end{flushright}
\smallskip
\begin{center} \LARGE
{\bf Twisting the Hagedorn temperature in planar $\NN=4$ super Yang-Mills}
 \\[12mm] \normalsize
{\bf  Simon Ekhammar${}^{a,b}$,  Joseph A. Minahan${}^{b,c,d}$, and Charles Thull${}^{e}$ } \\[8mm]
 {\small\it
    ${}^a$ Mathematics Department, Kings College London,\\
    The Strand, London WC2R 2LS, UK
     
     \smallskip
     \centerline{\it and}
     
  ${}^b$Department of Physics and Astronomy,
     Uppsala University,\\
     Box 516,
     SE-751\,20 Uppsala,
     Sweden
     
    \smallskip
     \centerline{\it and}
     
     \smallskip
     ${}^c$ Centre for Geometry and Physics, Uppsala University, Uppsala, Sweden
     
    \smallskip
     \centerline{\it and}
     
     \smallskip
     ${}^d$Center for Theoretical Physics -- a Leinweber Instititute,\\
     Massachusetts Institute of Technology\\
     Cambridge, MA 02139, USA

     \smallskip
     \centerline{\it and}
     
     \smallskip
     ${}^e$ Centre for Mathematics Innovation, City St George's, University of London,\\
     Northampton Square, EC1V 0HB London, UK
     
     }

  \medskip 
   \texttt{ \{simon.ekhammar, joseph.minahan\}@physics.uu.se, charles.thull@citystgeorges.ac.uk}

\end{center}
\vspace{7mm}
\begin{abstract}
We consider planar $\NN=4$ super Yang-Mills at finite temperature with chemical potentials that couple either to the $R$-charges or the spins of the operators.  We find expressions for the Hagedorn temperatures at both zero coupling by explicitly counting states, and at strong coupling using the string theory dual.  We then apply the quantum spectral curve (QSC) to this problem, which adds additional twists to the $Q$-functions.  For a single chemical potential $\mu$ coupled  to one of the $R$-charges, we find the analytic weak-coupling Hagedorn temperature to one-loop order for any value of $\mu$,  and to two-loop order for $\mu=1/2$. We then solve the QSC numerically, showing that at strong coupling there is good  agreement with the string theory prediction to order $1/\lambda^{1/4}$.  This provides further evidence for a recent conjecture of Harmark for the form of the world-sheet zero-point shift.  We also use the QSC to find the analytic one-loop correction to the Hagedorn temperature with non-zero chemical potentials coupled to the spins.
\end{abstract}

\eject
\normalsize

\tableofcontents

\section{Introduction}

The quantum spectral curve (QSC) is a powerful method to compute physical quantities in integrable gauge theories \cite{Gromov:2013pga,Gromov:2014caa}. In the case of $\NN=4$ super Yang-Mills (SYM) in the planar limit, one can use the QSC to interpolate between  weak 't Hooft coupling, $\lambda=g_{YM}^2N$,  where perturbation theory is relevant, to strong coupling, where one can compare with predictions from the dual string theory.  In fact, the quantum spectral curve {\it is} the best available tool for doing perturbative calculations in planar $\NN=4$, where twenty orders in perturbation theory are often reached \cite{Marboe:2018ugv}.  

Unfortunately, at strong coupling it is only known how to extract numerical results from the QSC~\cite{Gromov:2023hzc,Ekhammar:2024rfj}. Nevertheless, using some partial results and good guesses, it is often possible to find the analytic coefficients for the perturbative expansion in terms of  $1/\sqrt{\lambda}$, even though deriving the same results from a direct string world-sheet computation seems impossible with the present limitations.  On the other hand, the QSC can provide clues on what to look for in the string world-sheet computation, so it is fruitful to use it in order to investigate various scenarios.

One such application uses the QSC to find the Hagedorn temperature, $T_H$, for planar $\NN=4$ SYM on $S^3$ as a function of $\lambda$.  The value of $T_H$ at $\lambda=0$ was first computed by Sundborg~\cite{Sundborg:1999ue}, where he further showed that $T_H$ equals the de-confinement temperature.  This was independently shown in \cite{Aharony:2003sx}.  The first perturbative correction to $T_H$ was found in \cite{Spradlin:2004pp}, but to go beyond one should exploit the underlying integrability of the planar gauge theory.  

This was first done in a remarkable set of papers by Harmark and Wilhelm \cite{Harmark:2017yrv,Harmark:2018red,Harmark:2021qma}, which culminated in the seven-loop perturbative correction to $T_H$, as well as a numerical result for $T_H$ valid for all values of the coupling which matches on to the perturbative result.  At strong coupling they showed that $T_H$ scales as $\lambda^{1/4}$ with a coefficient that is consistent with the AdS/CFT correspondence \cite{Harmark:2021qma}.  Furthermore, their numerical result for the first sub-leading coefficient was later shown to be consistent with the leading correction coming from a winding string in thermal AdS \cite{Maldacena_unpub,Urbach:2022xzw}.

Subsequently, the authors of this paper improved the numerics in the QSC calculation to conjecture the next two terms in the strong coupling expansion \cite{Ekhammar:2023glu}.  At second order the coefficient depends on a non-trivial contribution  to the world-sheet zero-point shift.  The form of this shift was assumed to be similar to a corresponding shift found when computing $T_H$ in the plane-wave background, which is known exactly.  Furthermore, the form taken by the shift did not affect the third-order coefficient, which was consistent with the numerical results.  Finally, the calculation of $T_H$ was also done for ABJM \cite{Aharony:2008ug} using the corresponding QSC, where the form of the zero-point shift was shown to be consistent with the leading terms in the strong coupling expansion \cite{Ekhammar:2023cuj}.

While these results are encouraging, the implementation of the zero-point correction was somewhat {\it ad hoc.}  Recently, Harmark put this on firmer ground by arguing that the correction to the zero-point shift must be covariant with an overall coefficient \cite{Harmark:2024ioq}.  This coefficient is then determined from $T_H$ of the plane-wave limit and leads to the conjectured form for the second-order correction in \cite{Ekhammar:2023glu,Ekhammar:2023cuj} \footnote{For another argument that leads to the conjectured form see \cite{Bigazzi:2023hxt}.}.  However, this covariant term adds an additional term to the third-order correction that needs to be canceled \cite{Harmark:2024ioq}.  Harmark found a term that does the trick, but this term is zero for the plane-wave, hence the overall coefficient cannot be directly determined from it.  

It would be useful to have other situations where one can test the conjectures in \cite{Harmark:2024ioq} for the strong-coupling regime.  One way to do this is to add chemical potentials that couple to the $R$-charges or the spins of the operators.  From the QSC point of view, this corresponds to putting additional twists on the $Q$-functions \cite{Harmark:2021qma}.  One particular case that will play a large role in this paper is where there is one nonzero chemical potential $\mu$ that couples to a $U(1)$ subgroup of the $SO(6)$ $R$-symmetry group.  The partition function then takes the form
\be\label{eq:partitionFunction}
\ZZ=\sum_{\rm states} e^{-(\Delta-Q\mu)/T}\,,
\ee
where BPS-states with $\Delta=Q$ dominate \eqref{eq:partitionFunction} as $\mu$ approaches 1. Consequently, the Hagedorn temperature approaches zero in this limit for any value of $\lambda$.  For $\mu>1$ the partition function is clearly divergent and a Hagedorn temperature does not exist. For this same reason, if we consider states with $\Delta\sim N^2$, then this divergence is directly linked to the instability of black holes with $\mu>1$ \cite{Choi:2024xnv}. A similar partition function with an imaginary chemical potential coupled to a diagonal $U(1)$ subgroup of the $SO(6)$ $R$-symmetry was used in~\cite{Cherman:2020zea} to study the temperature of (partial) confinement transitions in $\NN=4$ SYM at large $N$.

The Hagedorn temperature as a function of $\mu$, $T_H(\mu)$, is found by assuming that  \eqref{eq:partitionFunction} is dominated by products of single trace operators.  At $\lambda=0$  $T_H(\mu)$ satisfies a known equation, where one can expand about either small $\mu$ or $\mu\to1$  \cite{Yamada:2006rx,Harmark:2006di}.  In  \cite{Harmark:2006di} it was also shown how to generalize the procedure in \cite{Spradlin:2004pp} to find a functional equation for the one-loop correction to $T_H(\mu)$. 

In this paper we will use the AdS/CFT correspondence to make predictions for $T_H(\mu)$, up to order $1/\lambda^{1/2}$ in the strong-coupling expansion.  We make use of the conjectures in \cite{Harmark:2024ioq} to find the expansion coefficients at order $1/\lambda^{1/4}$ and $1/\lambda^{1/2}$.  Using the QSC we then connect the weak- to the strong-coupling behavior  numerically, where we show that the $1/\lambda^{1/4}$ coefficient is within the error bars of the prediction.  Unfortunately, the numerics are currently not stable enough  to meaningfully compare with the prediction for the $1/\lambda^{1/2}$ coefficient.

One can also consider twists of the spins, where the partition function takes the form
\begin{eqnarray}
    \ZZ=\sum_{\rm states} e^{-(\Delta-\Omega_1 S_1-\Omega_2S_2)/T}\,,
\end{eqnarray}
where $S_1$ and $S_2$ are the two $SO(4)$ spins and $\Omega_1$ and $\Omega_2$ are the corresponding chemical potentials. The case for one chemical potential but for a general $d$ 
conformal field theory was recently considered in \cite{Seitz:2025wpc} where they found the leading correction to the flat space limit for the Hagedorn temperature.  Here, we make predictions for the next two coefficients in the $1/\lambda$ expansion for $T_H(\Omega_1,\Omega_2)$ using the criteria in \cite{Harmark:2024ioq}.  However, the numerics for the QSC are much less stable than for the $R$-charge chemical potential, so we are not yet able to verify the coefficients.

The rest of the paper is structured as follows:  In section \ref{sec:ZeroCoupling} we review known results and also present new results at weak-coupling for $T_H(\mu)$ and $T_H(\Omega_1,\Omega_2)$.  In section \ref{sec:ThermalScalar} we compute $T_H(\mu)$ and $T_H(\Omega_1,\Omega_2)$ at strong-coupling using the string theory dual.  In section \ref{sec:QSC} we describe the QSC and the results derived from it for $T_H(\mu)$. In section \ref{sec:Conclusion} we present some brief conclusions.

In a pair of ancillary files we provide the numerical QSC data presented in section~\ref{sec:QSC} and a Mathematica notebook to visualize them.

\section{The twisted Hagedorn temperature at weak coupling}\label{sec:ZeroCoupling}

This section follows the paper by Sundborg \cite{Sundborg:1999ue}, modified to account for one or more nonzero chemical potentials.  The case for nonzero chemical potentials coupled to the $R$-charges was first considered in \cite{Yamada:2006rx,Harmark:2006di}.  

We  consider  $\NN=4$ SYM on $R\times S^3$, where we set the radius of the $S^3$ to unity. Under the conformal transformation $R^4\to R\times S^3$ the dilation operator $\Delta$ maps to the Hamiltonian $H$, while local operators $\OO$ map to states in the  Hilbert space on $S^3$.  In the large-$N$ limit we need only consider single-trace operators, where at zero 't Hooft coupling their dimensions equal their bare dimensions.

The single trace operators have the form  $\Tr(\Phi_{j_1}\Phi_{j_2}\dots \Phi_{j_L})$, where the fields $\Phi_j$ transform in the adjoint representation of $SU(N)$.  The choices for the $\Phi_j$ are: 1) the three complex scalar fields $Z=\frac{1}{\sqrt{2}}(\phi^1+\io\phi^2)$, $X=\frac{1}{\sqrt{2}}(\phi^3+\io\phi^4)$ and $Y=\frac{1}{\sqrt{2}}(\phi^5+\io\phi^6)$, their complex conjugates, and covariant derivatives acting on them; 2) the fermion fields $\psi^a$ or $\bar\psi_a$, $a=1,\dots 4$, also with covariant derivatives; 3) the field strengths $F_{\mu\nu}$ and their covariant derivatives.  The $\phi^I$ have dimension $\Delta=1$, the $\psi^a$ and $\bar\psi_a$ have dimension $\Delta=3/2$, and the $F_{\mu\nu}$ have dimension $\Delta=2$.  Each derivative acting on the fields increases the dimension by $1$.  Some combinations of derivatives acting on the fields are zero and hence do not contribute to the states.  These are $\partial^2\phi^I=\dsl \psi^a=\dsl\bar\psi_a=\partial^\mu F_{\mu\nu}=0$ by the  equations of motion, and $\partial^\mu*F_{\mu\nu}=0$ by the Bianchi identity. We first focus on a $U(1)$ subgroup of the $SO(6)$ $R$-charge, where the $Z$ field has charge $+1$, $\bar Z$ has charge $-1$, $\psi^a$ has charge $+1/2$ and $\bar\psi_a$ has charge $-1/2$.  All other fields have charge $0$.

\subsection{$R$-charge twist at zero coupling}\label{sec:ZeroCouplingmu}
We now assume the system is at a temperature $T$ with a chemical potential $\mu$ for the above $U(1)$ $R$-charge \footnote{It is straightforward to generalize this to all three $R$-charges \cite{Yamada:2006rx,Harmark:2006di}, but this will not be necessary for our purposes.}.  Therefore, the partition function is
\be
\ZZ=\sum_{\rm states} e^{-(\Delta-Q\mu)/T}=\exp(\ZZ_1)
\ee
where $Q$ is the $U(1)$ $R$-charge and $\ZZ_1$ is the partition function for single trace states.  To count the single-trace states we  treat the different fields inside the trace as ``beads on a necklace".    If we define $y=e^{-\frac{1}{2T}}$ and the fugacity $z=e^{\frac{\mu}{2T}}$, then the scalar partition function is \cite{Yamada:2006rx,Harmark:2006di}
\be\label{Zphi}
Z_\phi(y,z)=\frac{(z^2+z^{-2}+4)y^2}{(1-y^2)^4}-\frac{(z^2+z^{-2}+4)y^6}{(1-y^2)^4}=\frac{(z^2+z^{-2}+4)y^2(1+y^2)}{(1-y^2)^3}\,,
\ee
where the denominator in the middle terms accounts for the derivatives in the four possible directions while the second term is to remove terms that are zero by the equations of motion.  The partition function for the fermionic beads is  \cite{Yamada:2006rx,Harmark:2006di}
\be
Z_{\psi}(y,z)=\frac{8(z+z^{-1})y^3}{(1-y^2)^4}-\frac{8(z+z^{-1})y^5}{(1-y^2)^4}=\frac{8(z+z^{-1})y^3}{(1-y^2)^3}\,,
\ee
where again the second term in the middle expression removes terms that are zero by the equations of motion.  Finally, the partition function for the field strength bead is
\be
Z_F(y)=\frac{6y^4}{(1-y^2)^4}-\frac{8y^6}{(1-y^2)^4}+\frac{2y^8}{(1-y^2)^4}=\frac{2y^4(3-y^2)}{(1-y^2)^3}\,,
\ee
where the second term in the middle expression removes terms that are zero by the equations of motion and the Bianchi identities, while the third term corrects for an over-subtraction.  Hence, the partition function for each bead is
\be\label{bead}
Z_{bead}(y,z)&=&Z_\phi(y,z)+Z_{\psi}(y,z)+Z_F(y)\nn\\
&=&\frac{(4+z^2+z^{-2})y^2+8(z+z^{-1})y^3+(10+z^2+z^{-2})y^4-2y^6}{(1-y^2)^3}\,.
\ee

\begin{figure}
    \centering
    \includegraphics[width=0.5\linewidth]{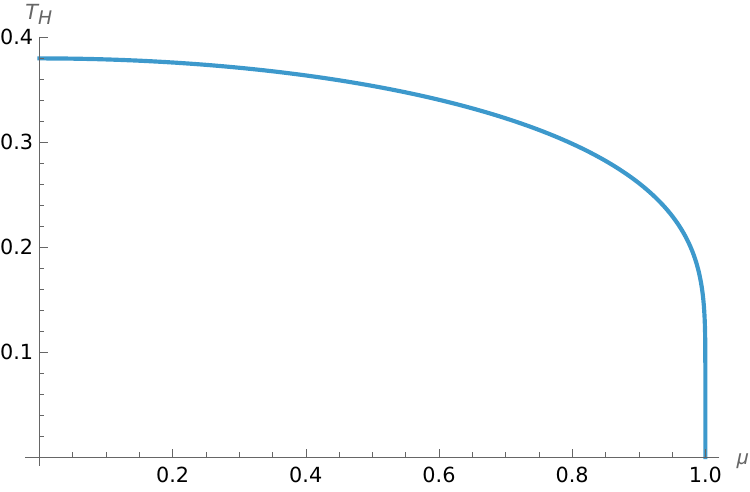}
    \caption{The Hagedorn temperature at zero coupling as a function of the chemical potential at zero coupling \cite{Harmark:2006di}.}
    \label{fig:ZeroCoupling}
\end{figure}

The Hagedorn temperature is determined by the behavior of long operators, in which case we can approximate the partition function as
\be
\ZZ_1\approx\sum_{L=2}^\infty\frac{1}{L}\left(Z_{bead}(y,z))\right)^{L}=-Z_{bead}(y,z)-\log\left(1-Z_{bead}(y,z)\right)\,,
\ee
which diverges when $Z_{bead}(y,z)=1$.  Solving this equation we end up with the relation
\be \label{eq:ZeroTempR}
\cosh\left(\frac{1}{2T_H}\right)\left(\cosh\left(\frac{1}{2T_H}\right)-\cosh\left(\frac{\mu}{2T_H}\right)\right)=2\,.\label{eq:beads:g0Result}
\ee
For $\mu=0$ this gives the solution $T_H=\frac{1}{2\log(2+\sqrt{3})}$ \cite{Sundborg:1999ue}, while in the limit $\mu\to 1$ we find that
\be
T_H\approx\left(\log\frac{1}{1-\mu}\right)^{-1}\,.
\ee The Hagedorn temperature as a function of $\mu$ over the domain $0\le\mu\le1$ is shown in Figure \ref{fig:ZeroCoupling}.

\subsection{$R$-charge twist at higher loops}

\subsubsection{$R$-charge twist at one loop}

Let us now turn to the case of the 1-loop computation of the Hagedorn temperature. In order to compute this quantity, we follow Spradlin-Volovich \cite{Spradlin:2004pp}. Expanding $y=y_{H}+g^2 \delta y_H+\dots$ with $y_H=\exp\left(-\frac{1}{2T_H^{(0)}}\right)$ and $T_{H}^{(0)}$ solving \eqref{eq:beads:g0Result}, the first correction is fixed as
\begin{eqnarray}\label{eq:yH1Loop}
    \delta y_{H} = -2\log y_H\frac{\expval{D_2(y)}}{\partial_{y}Z_{\text{bead}}(y)}\bigg|_{y=y_H}\,,
\end{eqnarray}
where $\expval{D_2(y)}$ is the expectation value of the $\mathcal{N}=4$ one loop dilation operator which is known exactly. We will not need the full expression for $D_{2}$, only the expectation value which takes the form
\begin{equation}\label{eq:1Loop}
    \expval{D_2(x)} = \sum_{j=0}^{\infty} h(j)\frac{V_j(y)}{(1-y)^4}\,
\end{equation}
where $h(j)$ is the harmonic number and $V_j$ are the characters of the module $\mathcal{B}^{\frac{1}{4},\frac{1}{4}}_{[1,0,1]}$ for $j=1$ and $\mathcal{C}^{1,1}_{[0,0,0],(\frac{j}{2}-1,\frac{j}{2}-1)^{*}}$ for $j\geq 2$, with $*$ prescribing a subtraction of all gauge-degrees of freedom, these characters are for example collected in \cite{Dolan:2002zh},\cite{Bianchi:2003wx}.

\begin{figure}
    \centering
    \includegraphics[width=0.5\linewidth]{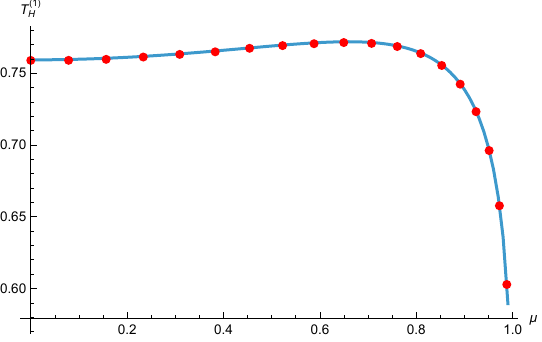}
    \caption{We display the 1-loop correction to the Hagedorn temperature for $\mu\in [0,1]$. The solid blue line is the exact analytic expression \eqref{eq:TH1Analytic} while the red points are from fitting the numerical QSC data shown in \autoref{fig:weakCoupling}.}
    \label{fig:placeholder}
\end{figure}

It is straightforward to do all sums, and in our particular case, we find
\begin{equation}
\begin{split}
    \expval{D_2(y,z)} &= -\frac{y^2 (y+z)^3 (y z+1)^3 \left(2 y^4 z-2 y^3 \left(z^2+1\right)-13 y^2 z-3 y
   \left(z^2+1\right)+z\right)}{\left(y^2-1\right)^6 z^4} \\
   &\quad \quad \quad -\left(Z_{\text{bead}}-1\right)^2\log(1-y^2)\,.
\end{split}
\end{equation}
Plugging this into \eqref{eq:yH1Loop}, re-expressing everything in terms of $T_H = T^{(0)}_H + g^2 T^{(1)}_H+\dots$ and finally simplifying using \eqref{eq:beads:g0Result} we obtain
\begin{equation}\label{eq:TH1Analytic}
T^{(1)}_{H} = T\frac{32 \tanh ^2\left(\frac{1}{2 T}\right)}{-\mu \sqrt{2} 
   \left(\cosh \left(\frac{1}{T}\right)-7\right)^{1/2} \cosh \left(\frac{1}{2
   T}\right)+\cosh \left(\frac{1}{T}\right)+5}\bigg|_{T=T_H^{(0)}}\,.
\end{equation}
One quickly verifies that indeed $T_H^{(1)}\big|_{\mu=0} = 2\,T_H^{(0)}\big|_{\mu=0}$ as it should be \cite{Spradlin:2004pp}. We compared this prediction to the numerical QSC results to be presented below, finding excellent agreement as shown in Figure~\ref{fig:placeholder}. 

\subsubsection{Higher loops}
In order to find the higher loop contributions to the Hagedorn temperature one is practically forced to use the perturbative QSC. In \cite{Harmark:2021qma} this was accomplished up to seven loops for the undeformed Hagedorn temperature. In principle, one can use the same methods to solve the $R$-twisted QSC. However, in practice the algorithm is significantly slower if the parameter $\mu$ is kept arbitrary. This is because the intermediate expressions become much more involved and hence require more computer time to simplify. This is a problem already encountered for $\beta$ and $\gamma$ deformations in \cite{Marboe:2019wyc,Levkovich-Maslyuk:2020rlp}.

To simplify the problem we set $\mu=\frac{1}{2}$, in which case $\sqrt{z}$ satisfies a simple polynomial equation as a consequence of \eqref{eq:ZeroTempR}. Solving the QSC perturbatively we reproduce the tree-level and one-loop results and furthermore find
\begin{align}
    T^{(2)}_{H}\bigg|_{\mu=\frac{1}{2}} =&\, T\,\log\left(1-y^{2}\right)\left(\frac{2 \left(44 y^{5/2}+104 y^{3/2}-4 y^2-57 y+47 \sqrt{y}-21\right)}{y^{5/2}}\right)\nonumber \\
    &+\frac{123475 y^{5/2}+281810 y^{3/2}-12601 y^2-159501 y+117087 \sqrt{y}-55652}{1352 y^{5/2}} \nonumber\\
     &+T\left(\scalebox{1.2}{$\frac{\left(-68 y^{3/2}+51 y^2+5 y+9 \sqrt{y}+1\right) \left(-3 y^{5/2}-8 y^{3/2}+y^3+5 y^2+5
   y-3 \sqrt{y}+1\right)}{169 y^5}$}\right)\bigg|_{T=T_H^{(0)}}\nonumber\\
    \simeq& -4.5913859014992799903\dots\,.
\end{align}
This result matches our numerical QSC results up to an error of $10^{-19}$, \textit{cf}. the ancillary Mathematica notebook and data file.

\subsection{Generic spin twists at zero coupling}\label{sec:spinTwistWeak}

\begin{figure}
    \centering
    \includegraphics[width=0.75\linewidth]{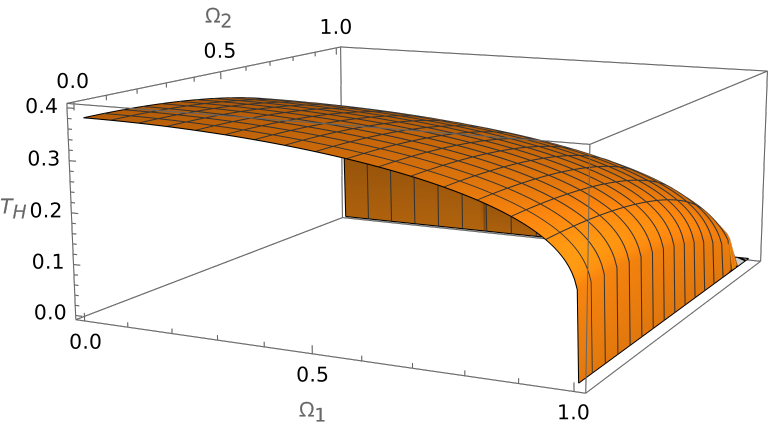}
    \caption{The Hagedorn temperature at zero coupling in $\N=4$ SYM against both chemical potentials for the spin.}
    \label{fig:Spin:BothChemPotentials}
\end{figure}

We next assume that we have chemical potentials $\Omega_1$ and $\Omega_2$ that couple to the two angular momenta on $S^3$,  $S_1$ and $S_2$.  The partition function then has the form
\begin{equation}
    \ZZ(\beta)=\sum_\textrm{states}e^{-\beta(\Delta-\Omega_1 S_1-\Omega_2 S_2)},
\end{equation} 
For $\lambda=0$ we can follow  the same route as in section \ref{sec:ZeroCouplingmu} to study its radius of convergence. The different contributions to the single bead partition function now take the form \begin{align}
    Z_\phi&=\frac{6y^2(1-y^4)}{(1-y^2z w)(1-y^2\frac{z}{w})(1-y^2\frac{w}{z})(1-\frac{y^2}{z w})},\\
    Z_\psi&=\frac{4y^3(1-y^2)(z+\frac{1}{z}+w+\frac{1}{w})}{(1-y^2z w)(1-y^2\frac{z}{w})(1-y^2\frac{w}{z})(1-\frac{y^2}{z w})},\\
    Z_F&=\frac{y^4(z^2+1+\frac{1}{z^2}+w^2+1+\frac{1}{w^2})-2y^6(z w+\frac{z}{w}+\frac{w}{z}+\frac{1}{z w})+2y^8}{(1-y^2z w)(1-y^2\frac{z}{w})(1-y^2\frac{w}{z})(1-\frac{y^2}{z w})}\,,
\end{align} 
where $y$ is the same as in \eqref{Zphi}, and  the fugacities are $z=e^{\frac{\Omega_1+\Omega_2}{2T}}$ and $w=e^{\frac{\Omega_1-\Omega_2}{2T}}$. As before, the Hagedorn temperature is obtained by setting the single bead partition function $Z_\textrm{bead}=Z_\phi+Z_\psi+Z_F$ equal to 1. We then get the following equation \begin{multline}
    \cosh \left(\frac{1}{2T_H}\right) \left(\cosh \left(\frac{1}{T_H}\right)-\cosh \left(\frac{\text{$\Omega_1$}}{T_H}\right)-\cosh \left(\frac{\text{$\Omega_2$}}{T_H}\right)-3\right)\\
    =4  \cosh \left(\frac{\text{$\Omega_1$}}{2 T_H}\right) \cosh \left(\frac{\text{$\Omega_2$}}{2
   T_H}\right).
\end{multline} We show the solution for $0\leq \Omega_1,\Omega_2\leq1$ in figure \ref{fig:Spin:BothChemPotentials}. Note that the Hagedorn temperature goes to zero if either $\Omega_1$ or $\Omega_2$ approaches 1.

\subsection{Spin twists at one loop}
We can use the same methods as in the case of the $R$-twist. While it is straightforward to allow for both $\Omega_{1}$ and $\Omega_{2}$ to be non-zero, the resulting expressions are very messy. We give an explicit expression for this general case in Appendix~\ref{app:OneLoopTwisted}. Here we focus instead on the case $\Omega_2 = 0$ where the expressions simplify significantly. We obtain
\begin{multline}\label{eq:Twist1Loop}
    \frac{T^{(1)}_{H}}{T_H^{(0)}} = -\frac{2}{\mathcal{D}}\bigg((y z-1)^2 (y+z)^6 \log \left(1-\frac{y^2}{z^2}\right)+(y-z)^2 (y z+1)^6 \log \left(1-y^2 z^2\right)\\
    -2 (y-z)(y z-1) (y z+1)^3 (y+z)^3 \log \left(1-y^2\right) \bigg)\bigg|_{T=T_{H}^{(0)}},
\end{multline}
where 
\begin{equation}
    \mathcal{D} =y^2 \left(z^2-1\right)^2 \left( \Omega_1 y \left(2 \left(y^2-1\right) z \left(z^2-1\right)+y
   \left(-\left(z^2 \left(-3 y^2+z^2+4\right)\right)-1\right)\right)+3 z^2\right).
\end{equation}

Once again QSC techniques can in principle be used to compute perturbative corrections to the Hagedorn temperature, but for general $\Omega_1,\Omega_2$ computations are slow due to complicated intermediate expressions. We therefore restrict ourselves to the case $\Omega_1=\frac{1}{2}$ and $\Omega_2=0$. Perturbatively solving the QSC we then find the following
\begin{align}
    &T^{(1)}_{H}\bigg|_{\Omega_{1}=\frac{1}{2},\Omega_{2}=0}\\
    &= \frac{\left(41 y^{5/2}+169 y^{3/2}-111 y^2-280 y+104 \sqrt{y}-88\right)}{13 \log (y)} \left(\Li_1(y) -2 \Li_1(y^2)+\Li_1(y^3)\right)\bigg|_{T=T_{H}^{(0)}}\nonumber
   \end{align}
which perfectly reproduces \eqref{eq:Twist1Loop} specialized to $z=\frac{1}{\sqrt{y}}$.

\section{The twisted Hagedorn temperature at strong coupling}\label{sec:ThermalScalar}

\subsection{$R$-charge twist at strong coupling}

We next consider the Hagedorn temperature at strong coupling in the presence of a chemical potential for the $R$-charge.  In this case the chemical potential leads to a background gauge field in the bulk with the full 10-dimensional metric given by \cite{Hawking:1999dp}
\be
ds^2=(1+R^2)d\tau^2+\frac{dR^2}{1+R^2}+R^2 d\Omega_{3}^2+(1-Z^2)(A_\mu dx^\mu+d\psi)^2+\frac{dZ^2}{1-Z^2}+Z^2 d\Omega_3^2\,.
\ee
Here we have set the radii of the $AdS_5$ and $S^5$ to 1 and make the identification  $\tau\equiv\tau+\beta$.  In particular, for a chemical potential $\mu$ the gauge field is given by $A_\mu dx^\mu=i\,\mu d\tau$. The world-sheet fermions have anti-periodic boundary conditions which shifts the zero-point energy of the string to a nonzero value, $C=C_0+\Delta C$ where $C_0=-2/\al'$ is the shift in flat space.

The thermal scalar corresponds to a string that winds once around the $\tau$ direction. If $\beta$ is tuned appropriately, then this winding mode is very light and we can use the supergravity approximation.  The winding mode is a scalar field $\chi$ and its contribution to the action is
\be 
\int d^{d+1}X \sqrt{g}\left(\nabla^\mu\chi\nabla_\mu\chi+m^2(R,Z)\chi^2\right)\,,
\ee
where $m^2(R,Z)$ is the radial dependent mass-function
\be
m^2(R,Z)=(1+R^2)\left(\frac{\beta}{2\pi\al'}\right)^2-(1-Z^2)\left(\frac{\beta\mu}{2\pi\al'}\right)^2+C\,.
\ee
For $\beta$ properly chosen, $\chi$ is massless and so is independent of $\tau$.  We also assume that $\chi$ depends only on $R$ and $Z$, which leads to the equation
\be\label{EOM}
&&-\frac{1}{2}\frac{1}{R^{3}}\frac{d}{dR} R^{3}\frac{d}{dR} \chi(R,Z)-\frac{1}{2}\frac{1}{Z^{3}}\frac{d}{dZ} Z^{3}\frac{d}{dZ} \chi(R,Z)+\frac{1}{2}\left(\frac{\beta}{2\pi\al'}\right)^2(R^2+\mu^2Z^2)\,\chi(R,Z)\nn\\
&&\qquad\qquad\qquad\qquad+\Delta H\,\chi(R,Z)=-\frac{1}{2}\left(C+\left(\frac{\beta}{2\pi\al'}\right)^2(1-\mu^2)\right)\chi(R,Z)\,,
\ee
where 
\be\label{Hpert}
\Delta H &=&\Delta H_1+\Delta H_2\ee
with
\be\label{Hpert12}
\Delta H_1=-\frac{1}{2}\frac{1}{R^{3}}\frac{d}{dR} R^{5}\frac{d}{dR}\,,&&\qquad \Delta H_2=\frac{1}{2}\frac{1}{Z^{3}}\frac{d}{dZ} Z^{5}\frac{d}{dZ}\,.
\ee
Clearly, the $R$ and $Z$ variables are separable.  

\subsubsection{$\mu\lesssim 1$}\label{muls1}

If we assume that $\mu\lesssim 1$,  then  $\chi(R,Z)$ falls off exponentially away from $Z=0$.  With this assumption, solving \eqref{EOM}  is equivalent to solving the radial Schr\"odinger equation for two perturbed four-dimensional harmonic oscillators with frequencies $\om_1=\frac{\beta}{2\pi\al'}$, $\om_2=\frac{\beta\,\mu}{2\pi\al'}$, and energy
\be\label{2HOenergy}
E=-\frac{1}{2}\left(C+\left(\frac{\beta}{2\pi\al'}\right)^2(1-\mu^2)\right)\,.
\ee

As argued in \cite{Harmark:2024ioq}, $\Delta C$ can depend on the coordinates.  In particular, it should have a covariant form, which to leading order is given by 
\be\label{DC}
\Delta C=\frac{\log2}{2\pi^2\al'}R_{\mu\nu}V^\mu V^\nu\,,
\ee
where $R_{\mu\nu}$ is the Ricci tensor and $V^\mu$ a  time-like Killing vector with $V^0=\beta$, in order to match the prediction for the Hagedorn temperature for the plane-wave solution, which is exact.  This was then shown to be consistent with the prediction in \cite{Ekhammar:2023glu} based on the numerical integrability calculations.

To lowest order we  set $C=-2/\al'$ and ignore $\Delta H$.  The harmonic oscillator ground-state energy is then
\be\label{Eeq0}
E=\frac{1}{2}\left(\frac{2}{\al'}-\left(\frac{\beta}{2\pi\al'}\right)^2(1-\mu^2)\right)=\frac{4}{2}(\om_1+\om_2)=\frac{4}{2}\,\frac{\beta}{2\pi\al'}(1+\mu)\,.
\ee
Solving for $\beta$ we get
\be
\beta=\frac{\pi\al'}{1-\mu}\left(\sqrt{\frac{8(1-\mu)}{\al'(1+\mu)}+16}-4\right)\,.
\ee
Inverting this, we find the  Hagedorn temperature to leading order is
\be
T_H=\frac{1}{\beta}=\frac{\sqrt{1-\mu^2}}{2\pi\sqrt{2\al'}}+\frac{1+\mu}{2\pi}+\dots\,.
\ee
The correction from flat-space with $\mu=0$ matches  \cite{Urbach:2022xzw}.

We now compute the next two orders in $\sqrt{\al'}$.    For this we need up to second order in perturbation theory for the perturbed oscillators, as well as the correction $\Delta C$ in \eqref{DC}.  Inserting $R_{00}=-4(1+R^2)-4\mu^2(1-Z^2)$ into \eqref{DC}, we have that $C$ is
\begin{equation}\label{zptsh}
    C=-\frac{2}{\alpha'}-\frac{\beta^2}{2\pi^2\alpha'}4\log(2)[(1+\mu^2)+(R^2-\mu^2Z^2)] +\OO(\al').
\end{equation}
In the $\mu=0$ case it was argued in \cite{Harmark:2024ioq} that the effect of the $R^2$ term is canceled out to the desired order by the inclusion of the kinetic term 
\be\label{extrakinterm}
\al'\log(2)R^{\mu\nu}\nabla_\mu\nabla_\nu\chi
\ee
on the left hand side of \eqref{EOM}.  It is straightforward to check that this same term will also eliminate the effect of the $Z^2$ term.  We will hence drop these terms in \eqref{zptsh} and assume the original kinetic term.
The first order correction to the energy cancels from the two oscillators and we are left with the second order term
\be
\Delta E=-\frac{3}{4}\left(\frac{1}{\om_1}+\frac{1}{\om_2}\right)=-\frac{3\pi\al'}{2\beta\mu}(1+\mu)\,.
\ee
Therefore,   \eqref{Eeq0} becomes
\be\label{Eeq1}
&&    \frac{1}{2}\left(\frac{2}{\al'}-\left(\frac{\beta}{2\pi\al'}\right)^2(1-\mu^2)+\frac{\beta^2}{2\pi^2\alpha'}4\log(2)(1+\mu^2)\right)\nn\\
&&\qquad\qquad\qquad\qquad\qquad\qquad=2\frac{\beta}{2\pi\alpha'}(1+\mu)-\frac{3\pi\al'}{2\beta\mu}(1+\mu)\,.
\ee
This can be solved in a series expansion in $\al'$,
and gives the Hagedorn temperature
\be
    T_H&=&\frac{\sqrt{1-\mu^2}}{2\pi\sqrt{2\alpha'}}+\frac{1+\mu}{2\pi}+\frac{(1+\mu)^2-(1+\mu^2)4\log(2)}{2\pi\sqrt{2(1-\mu^2)}}\sqrt{\al'}-\frac{3(1-\mu^2)(1+\mu)}{32\pi\mu}\al'\nn\\
    &&\qquad\qquad\qquad\qquad\qquad\qquad+{\rm O}((\al')^{3/2}).
\ee
Using the AdS/CFT dictionary, $\al'=\frac{1}{\sqrt{\lambda}}$, we find
\be
T_H&=&\frac{\sqrt{(1-\mu^2)}}{\sqrt{2\pi}}g^{1/2}+\frac{1+\mu}{2\pi}+\frac{(1+\mu)^2-(1+\mu^2)4\log(2)}{4\sqrt{2\pi^3(1-\mu^2)}}g^{-1/2}\nn\\
    &&\qquad\qquad\qquad\qquad\qquad\qquad-\frac{3(1-\mu^2)(1+\mu)}{128\pi^2\mu}g^{-1}+{\rm O}((g)^{-3/2})    \,,\label{THd}
\ee
where we have defined 
\be\label{coupling} 
g\equiv \frac{\sqrt{\lambda}}{4\pi}\,.
\ee
It is interesting that this expression looks  simpler than the $\mu=0$ case.  If we let $\mu\to1$, then we should consider the effective coupling $\tilde g=g(1-\mu^2)$ \footnote{In \cite{Harmark:2006di,Harmark:2006ta} the authors considered a different rescaling, where $\tilde T=\frac{T}{1-\mu}$ and $\tilde\lambda=\frac{\lambda}{1-\mu}$, which is appropriate for two $R$-charge chemical potentials $\mu_1=\mu_2=\mu\to1$, and weak 't Hooft coupling.}, in which case we find
\be\label{THdeff}
    T_H&=&\frac{1}{\sqrt{2\pi}}\tilde g^{1/2}+\frac{1}{\pi}+\frac{1-2\log(2)}{\sqrt{2\pi^3}}\tilde g^{-1/2}
+{\rm O}((\tilde g)^{-3/2})   \,.
\ee

\subsubsection{$\mu\lesssim\sqrt{\al'}\ll1$}
When $\mu$ is very small the harmonic oscillator approximation breaks down for the wave-function on $S^5$.  In this case one should instead write $\chi(R,Z)=\psi(R)Y(Z)$, where 
\be\label{YZ}
Y(Z)=\sum_{n=0}^\infty a_n S_{2n}(Z)
\ee
and the $S_{n}(Z)$ are the order $n$ spherical harmonic polynomials on $S^5$, which satisfy
\be
-\frac{1}{Z^3}\frac{d}{dZ} (1-Z^2) Z^3\frac{d}{dZ} S_{n}(Z)=n(n+4)S_{n}(Z)\,.
\ee
The $S_n(Z)$ also satisfy the normalization condition
\be
\int_0^\infty dZ\,Z^3\,  S_n(Z) S_m(Z)=\delta_{nm}\,.
\ee
One can further show that
\be
Z^2 S_{2n}(Z)&=&\frac{\sqrt{(n+1)(n+2)}}{2(2n+3)}S_{2n+2}(Z)+\frac{2(n+1)^2}{(2n+1)(2n+3)}S_{2n}(Z)\nn\\
&&\qquad\qquad\qquad\qquad\qquad\qquad+\frac{\sqrt{n(n+1)}}{2(2n+1)}S_{2n-2}(Z)\,.
\ee
One can then solve for the $a_n$ in \eqref{YZ} order by order in $\mu^2$ such that $Y(Z)$ is the lowest eigenstate of $H_2$, where $H_2$ is given by
\be
H_2=-\frac{1}{2}\frac{1}{Z^3}\frac{d}{dZ} (1-Z^2) Z^3\frac{d}{dZ}+\frac{1}{2}\left(\frac{\mu\,\beta}{2\pi\al'}\right)^2Z^2\,.
\ee
It is straightforward to show  that $H_2Y(Z)=\E(\om_2)Y(Z)$, where
\be
\E(\om_2)=\frac{1}{3}\om_2^2-\frac{1}{432}\om_2^4-\frac{1}{38880}\om_2^6+\frac{11}{44789760}\om_2^8+\rm{O}(\om_2^{10})\,,
\ee
and $\om_2$ has the same definition as above \eqref{2HOenergy}.
This series is convergent if $\om_2=\left(\frac{\mu\,\beta}{2\pi\al'}\right)\lesssim 5.7$, hence if $\beta\sim \sqrt{\al'}$, then $\mu\lesssim\sqrt{\al'}$ for this regime to be valid.  Assuming this is the case, then  \eqref{Eeq1} is modified to
\be\label{Eeq2}
&&    \frac{1}{2}\left(\frac{2}{\al'}-\left(\frac{\beta}{2\pi\al'}\right)^2+\om_2^2+\frac{\beta^2}{2\pi^2\alpha'}4\log(2)\right)=2\frac{\beta}{2\pi\alpha'}+3-\frac{3\pi\al'}{2\beta}+\E(\om_2)\,,
\ee
where we dropped the subleading term from the chemical potential contributing to $\Delta C$.  Since $\E(\om_2)\lesssim1$, the chemical potential only affects the Hagedorn temperature to order $\sqrt{\al'}$. Hence, to find $T_H$ to order $\al'$, it is sufficient to replace $\beta$ in $\omega_2$ with $\beta_H^{(1)}=2\pi(\sqrt{2\al'}-2\al')$, which includes the first-order correction to the flat-space limit. This then leads to 
\be\label{THd2}
    T_H(\alpha')&\approx&\frac{1}{2\pi\sqrt{2\alpha'}}+\frac{1}{2\pi}+\frac{5+f(\tilde\mu)-8\log(2)}{4\sqrt{2}\pi}\sqrt{\al'}\nn\\
    &&\qquad\qquad\qquad+\frac{45+16f(\tilde\mu)+8\tilde\mu f'(\tilde\mu)}{32\pi}\al'+{\rm O}((\al')^{3/2})\,,
\ee
where $\tilde\mu\equiv \frac{\mu}{\sqrt{\al'}}$ is a rescaled chemical potential and
\be
f(\tilde\mu)=\E(\sqrt{2}\tilde\mu)-\tilde\mu^2=-\frac{1}{3}\tilde\mu^2-\frac{1}{108}\tilde\mu^4-\frac{1}{4860}\tilde\mu^6+\frac{11}{2799360}\tilde\mu^8+\rm{O}(\tilde\mu^{10})\,.
\ee
The series converges if $\tilde \mu\lesssim 4.02$.

\subsection{Spin twists at strong coupling}\label{sec:spiTwistStrong}

We next consider the supergravity dual with spin twists. The case with a single spin twist was recently considered in \cite{Seitz:2025wpc}.  We generalize their method to include a second twist as well as higher order corrections.  Since there is now no R-symmetry twist we will work solely on AdS$_5$. The metric with the twists has the form
\begin{multline}
ds^2=(1+R^2)d\tau^2+\frac{dR^2}{1+R^2}\\
+R^2 \left[(1-Y^2)(A^{(1)}_\mu dx^\mu+d\phi_1)^2+\frac{dY^2}{1-Y^2}+Y^2(A^{(2)}_\mu dx^\mu+d\phi_2)^2\right]\,,
\end{multline}
where $A^{(1)}_\mu dx^\mu=i\,\Omega_1d\tau$ and  $A^{(2)}_\mu dx^\mu=i\,\Omega_2d\tau$. We then choose the new variables ${x=R\sqrt{1-Y^2}}$ and ${y=RY}$, in which case the metric becomes
\begin{multline}
    ds^2=(1+x^2+y^2)d\tau^2+dx^2+dy^2-\frac{(xdx+ydy)^2}{1+x^2+y^2}\\
    +x^2(i\,\Omega_1d\tau+d\phi_1)^2+y^2(i\,\Omega_2d\tau+d\phi_2)^2\,.
\end{multline}
The equation of motion that follows is now
\begin{multline}\label{EOM2}
    -\frac{1}{2}\left(\frac{1}{x}\frac{d}{dx}x\frac{d}{dx}+\frac{1}{y}\frac{d}{dy}y\frac{d}{dy}\right)\chi(x,y)+\frac{1}{2}\left(\frac{\beta}{2\pi\al'}\right)^2\left(\left(1-\Omega_1^2\right)x^2+\left(1-\Omega_2^2\right)y^2\right)\chi(x,y)\\
    \qquad\qquad+\Delta H\chi(x,y)=-\frac{1}{2}\left(C+\left(\frac{\beta}{2\pi\al'}\right)^2\right)\chi(x,y)\,,
\end{multline}
where the perturbed Hamiltonian is
\begin{eqnarray}
    \Delta H=-\frac{1}{2}\left(x^2\frac{d^2}{dx^2}+y^2\frac{d^2}{dy^2}+2xy\frac{d}{dx}\frac{d}{dy}+5x\frac{d}{dx}+5y\frac{d}{dy}\right)-\Delta H_{kin}\,.
\end{eqnarray}
$\Delta H_{kin}$ is the contribution of the kinetic term in \eqref{extrakinterm}, which to the relevant order in $\alpha'$ is
\begin{equation}
    \Delta H_{kin}=-4\al'\log(2)\left(\frac{1}{x}\frac{d}{dx}x\frac{d}{dx}+\frac{1}{y}\frac{d}{dy}y\frac{d}{dy}\right)\,.
\end{equation}    
We also have that $R_{00}=-4[1+(1-\Omega_1^2)x^2+(1-\Omega_2^2)y^2]$, hence $C$ is given by
\begin{equation}\label{zptsh2}
    C\equiv C(x,y)=-\frac{2}{\alpha'}-\frac{\beta^2}{2\pi^2\alpha'}4\log(2)[1+(1-\Omega_1^2)x^2+(1-\Omega_2^2)y^2] +\OO(\al').
\end{equation}
Like the situation in section \ref{muls1}, the effect of $\Delta H_{kin}$ cancels off the contribution from the $x^2$ and $y^2$ terms in \eqref{zptsh2} to order $\OO(\al')$, so we drop $\Delta H_{kin}$ and replace $C$ with $C(0,0)$.

The equation in \eqref{EOM2} is the radial Schr\"odinger equation for two two-dimensional harmonic oscillators with frequencies $\om_i=\frac{\beta\sqrt{1-\Omega_i^2}}{2\pi\al'}$, along with a perturbative potential $\Delta H$.  The desired solution then satisfies
\begin{eqnarray}
    \om_1+\om_2+\Delta E^{(1)}+\Delta E^{(2)}=-\frac{1}{2}\left(C(0,0)+\left(\frac{\beta}{2\pi\al'}\right)^2\right)\,,
\end{eqnarray}
where $\Delta E^{(1)}$ and $\Delta E^{(2)}$ are the first and second order perturbative corrections, respectively. It is straightforward to show that
\begin{eqnarray}
    \Delta E^{(1)}&=&3\nn\\
    \Delta E^{(2)}&=&-\frac{1}{4}\left(\frac{1}{\om_1}+\frac{1}{\om_2}+\frac{2}{\om_1+\om_2}\right)\,.
\end{eqnarray}
Hence, we find the relation
\begin{eqnarray}
\label{Eeq3}
 && \frac{1}{2}\left(\frac{2}{\al'}-\left(\frac{\beta}{2\pi\al'}\right)^2+\frac{\beta^2}{2\pi^2\alpha'}4\log(2)\right)=\frac{\tilde\beta_1+\tilde\beta_2}{2\pi\alpha'}+3-\frac{\pi\al'}{2}\left(\frac{1}{\tilde\beta_1}+\frac{1}{\tilde\beta_2}+\frac{2}{\tilde\beta_1+\tilde\beta_2}\right)\nn\\
  &&\qquad\qquad\qquad\qquad\qquad\qquad\qquad\qquad\qquad\qquad\qquad\qquad\qquad+\OO(\al')\,,  
\end{eqnarray}
where $\tilde\beta_i\equiv\beta\sqrt{1-\Omega_i^2}$.  Solving \eqref{Eeq3} and using the AdS/CFT dictionary, we find
\begin{multline}
 T_H=\frac{1}{\sqrt{2\pi}}g^{1/2}+\frac{\sqrt{1-\Omega_1^2}+\sqrt{1-\Omega_2^2}}{4\pi}+\frac{6+\left(\sqrt{1-\Omega_1^2}+\sqrt{1-\Omega_2^2}\right)^2-16\log(2)}{16\sqrt{2\pi^3}}g^{-1/2}\\
    +\frac{1}{128\pi^2}\left(\frac{23-24\Omega_1^2}{\sqrt{1-\Omega_1^2}}+\frac{23-24\Omega_2^2}{\sqrt{1-\Omega_2^2}}-\frac{2}{\sqrt{1-\Omega_1^2}+\sqrt{1-\Omega_2^2}}\right)g^{-1}+{\rm O}((g)^{-3/2})    \,.\\\label{THdspin}  
\end{multline}
One can easily check that this matches the result in \cite{Ekhammar:2023cuj} when $\Omega_1=\Omega_2=0$. Also, taking $\Omega_1=\Omega$ and $\Omega_2=0$, the leading two terms reproduce the result of~\cite{Seitz:2025wpc}.

Unlike the case for the $R$-charge chemical potential in \eqref{THd}, \eqref{THdspin} smoothly connects to the $\Omega_1=\Omega_2=0$ result in \cite{Ekhammar:2023cuj}. One can also see that the perturbative result breaks down if $\Omega_1\to1$ (and/or $\Omega_2\to1$).  One can understand this by examining \eqref{EOM2} and \eqref{zptsh2}, where we see that the $x$  dependence cancels in the effective potential and the solution de-localizes in $x$ such that $\chi(x,y)=\chi(y)$.  The effect of this is to replace the right hand side of \eqref{Eeq3} with 
$$
\frac{\tilde\beta_2}{2\pi\alpha'}+2-\frac{3\pi\al'}{2\tilde\beta_2}.  
$$
One then finds the  Hagedorn temperature 
\begin{multline}
 T_H=\frac{1}{\sqrt{2\pi}}g^{1/2}+\frac{\sqrt{1-\Omega_2^2}}{4\pi} +\frac{5-\Omega_2^2-16\log(2)}{16\sqrt{2\pi^3}}g^{-1/2}\\+\frac{13-16\Omega_2^2}{128\pi^2\sqrt{1-\Omega_2^2}}g^{-1}+{\rm O}((g)^{-3/2})\,.  
\end{multline}
which is nonzero for the limiting chemical potential, differing from the zero-coupling behavior.
If we also send $\Omega_2\to1$, then the solution further de-localizes in the $y$ direction and the right hand side of \eqref{Eeq3} is set to zero.  The Hagedorn temperature then becomes
\begin{eqnarray}
 T_H&=&\frac{1}{\sqrt{2\pi}}g^{1/2}-\frac{\log(2)}{\sqrt{2\pi^3}}g^{-1/2}+{\rm O}((g)^{-3/2})    \,,\label{THdspinlim}  
\end{eqnarray}
  Notice that the form in \eqref{THdspinlim} is similar to that found in \eqref{THdeff} for the $R$-charge chemical potential using an effective coupling.

\section{The twisted quantum spectral curve}\label{sec:QSC}

The quantum spectral curve for $AdS_5\times S^5$ was initially developed in~\cite{Gromov:2013pga,Gromov:2014caa} to solve the spectral problem. We start this section by briefly reviewing its important aspects before discussing how, based on~\cite{Harmark:2021qma}, we adapt it for the Hagedorn temperature with one chemical potential for the R-symmetry.

The main objects of the $AdS_5$ QSC are the 256 $Q$-functions which all depend on a single complex variable $u$, the spectral parameter. Of these 256 functions only 24 are necessary for our  algorithm and  we call them  $\bP_a$-, $\bQ_i$- and $Q_{a|i}$-functions. Here $a$~is a fundamental $\SU(4)_R$ R-symmetry index and $i$ is a fundamental $\SU(2,2)$ conformal symmetry index\footnote{Note that in \cite{Harmark:2021qma} and \cite{Ekhammar:2023cuj} the $\bP$-functions carry $SU(2,2)$ indices while the $\bQ$ functions carry $SU(4)$ indices.  We have flipped their roles to match the standard convention in the QSC literature.}. The $Q$-functions are related by finite difference equations called $QQ$-relations \begin{align}
    Q_{a|i}^+-Q_{a|i}^-&=\bP_a\bQ_i, & \bQ_i&=-\bP^a Q_{a|i}^+,
\end{align} where we use the notation $f(u)^\pm=f(u\pm\frac{\io}{2})$ and $\bP^a=\chi^{ab}\bP_b$ with $\chi^{ab}=(-1)^a\delta_{a+b,5}$. In addition, the $Q$-functions satisfy the constraint $\det(Q_{a|i})=1$.

The $\bP$-functions have a single square-root branch cut along the interval $[-2g,2g]$ on the top $u$-plane Riemann sheet, where the parameter $g$ is the coupling defined in \eqref{coupling}. The $\bQ$-functions instead have a long square-root branch cut on ${(-\infty,-2g]\cup[2g,\infty)}$. Finally, the $Q_{a|i}$  have a tower of branch cuts in the lower half-plane ${[-2g-\io\frac{2n+1}{2},2g-\io\frac{2n+1}{2}]}$ with $ n\in\mathbb{Z}_{\geq 0}$.

The physical observable we compute enters the quantum spectral curve through the asymptotics of the $Q$-functions. The twisted quantum spectral curve for the Hagedorn temperature was first discussed in~\cite{Harmark:2021qma}. Specializing to the specific scenario with a single chemical potential for the R-symmetry discussed in section~\ref{sec:ZeroCoupling}, we find the large $u$ asymptotics,
\begin{align}
    \bP_a&\sim \begin{pmatrix}
        y^{-\io u\mu}\\
        u\, y^{-\io u\mu}\\
        y^{\io u\mu}\\
        u\, y^{\io u\mu}
    \end{pmatrix}, & 
    \bQ_i &\sim \begin{pmatrix}
        (-\frac{1}{y})^{\io u}\\
        u\, (-\frac{1}{y})^{\io u}\\
        (-\frac{1}{y})^{-\io u}\\
        u\,(-\frac{1}{y})^{-\io u}
    \end{pmatrix},
\end{align} where $y=\exp(-\frac{1}{2 T_H})$ contains the Hagedorn temperature and $(-1)^{\io u}=e^{\pi u}$. 

To close the QSC equations we supplement the $QQ$-relations with gluing conditions. As mentioned above, $\bQ_i$ should have on its first Riemann sheet only a long branch-cut on $(-\infty,-2g]\cup[2g,\infty)$. However, it is much easier to first construct another function, $\bQ_i^\downarrow$, which is analytic only in the upper half-plane. Then using the parity transformation, $u\rightarrow -u$, we get another function $\bQ_i^\uparrow$ which is analytic in the lower half-plane. $\bQ_i^\downarrow$ and $\bQ_i^\uparrow$ are then  ``glued'' along $u\in[-2g,2g]$ to give $\bQ_i$ with a long cut. Just like the case for  zero chemical potential~\cite{Harmark:2018red}, the gluing condition is
\begin{equation}\label{eq:gluing}
    \bQ_i^\downarrow(u+\io 0)=\textrm{diag}(e^{2\pi u},-e^{2\pi u},e^{-2\pi u},-e^{-2\pi u})_{i,j}\bQ_j^\downarrow(-u+\io 0)\,,
    \quad
    u\in (-2g,2g)\,.
\end{equation}

\subsection{The zero coupling solution}

At zero coupling all $Q$-functions have the form of an exponential multiplied by a polynomial. Thus, based on the large $u$ asymptotics, we make the {\it ansatz}
\begin{align}
	\bP_a&=\begin{pmatrix}
		e^{\frac{\io u \mu}{2T_H}}A_1^{(0)}\\
		e^{\frac{\io u \mu}{2T_H}}\bkt{A_2^{(0)}\io u+c_{2,0}^{(0)}}\\
		e^{-\frac{\io u \mu}{2T_H}}A_3^{(0)}\\
		e^{-\frac{\io u \mu}{2T_H}}\bkt{ A_4^{(0)}\io u+c_{4,0}^{(0)}}
	\end{pmatrix}, & \bQ_i&=\begin{pmatrix}
		e^{\bkt{\pi u+\frac{\io u }{2T_H}}}B_1\\
		e^{\bkt{\pi u+\frac{\io u }{2T_H}}}(\io u B_2+d_{2,0}^{(0)})\\
		e^{-\bkt{\pi u+\frac{\io u }{2T_H}}}B_3\\
		e^{-\bkt{\pi u+\frac{\io u }{2T_H}}}(\io u B_4+d_{4,0}^{(0)})
	\end{pmatrix}.
\end{align}
With this it is straightforward to solve the $QQ$-relations. To fix the gauge freedom, we choose the leading coefficients of the $\bP$- and $\bQ$-functions to be
\begin{align}
	A_1&=A_3=A_2=A_4=\frac{(1+y_0^{1-\mu})(1+y_0^{-1-\mu})}{(-1+y_0^{-2\mu})},\nonumber\\
	B_3&=B_1=1,\label{eq:gaugeChoice}\\
	B_2&=B_4=\frac{(1+y_0^{-1+\mu})^2(1+y_0^{-1-\mu})^2}{(-1+y_0^{-2})^2},\nonumber
\end{align} 
with $y_0=\exp(-\frac{1}{2T_H^{(0)}})$ and $T_H^{(0)}$ the Hagedorn temperature at $g=0$. The leading coefficients of the $Q_{a|i}$ are then fixed and take the form 
\begin{align}
	b_{a|i,0}&=-\wt{t}_i A_a B_i \frac{1}{\sqrt{y_0}^{\wt{t}_i+\mu}+\sqrt{y_0}^{-\wt{t}_i-\mu}}, & &\textrm{for}\ a=1,2,\ \text{and}\ i=1,2,3,4,\label{eq:QaiLeadCoeffs}
\end{align} with $\wt{t}_i=(1,1,-1-,1)$. The remaining coefficients are given by the relations in \eqref{eq:app:symQai}. To fix the remaining gauge freedom we choose the parity relations \begin{align}
    \bP_a(-u)&=(-1)^{a-1}\bP_{a+2}(u), & \bQ_i(-u)=(-1)^{i-1}\bQ_{i+2}(u),
\end{align} for $a,i=1,2$, where the second relation is only valid at zero coupling. We can then impose the gluing conditions to find the equation
\begin{eqnarray}\label{0couptw}
    y_0^2+\left(y_0^2+1\right) y_0^{-2 \mu }-\left(\left(y_0^4-6 y_0^2+1\right) y_0^{-\mu
   -1}\right)+1 = 0\,.
\end{eqnarray}
It is easy to check that \eqref{0couptw} is equivalent to \eqref{eq:beads:g0Result} and  admits a real solution only if $\mu\leq1$.

\subsection{The numerical solution of the QSC}

The numerical algorithm we will employ to solve the QSC was introduced in \cite{Gromov:2015wca}.  
It is by now a standard method and we relegate the details to appendix \ref{app:NumAlgorithm}.
The basic idea is to formulate the quantum spectral curve as a minimization problem of the gluing equation~\eqref{eq:gluing}.

With an implementation of this algorithm in Mathematica we would expect that computing the Hagedorn temperature for one value of the chemical potential would take up to one month, even with the use of a cluster. Thus, we have chosen to follow the example of \cite{Gromov:2023hzc} and instead implement the algorithm in C++. While our implementation was inspired by the code of~\cite{Gromov:2023hzc} and is also based on the CLN (\href{https://www.ginac.de/CLN/}{Class Library for Numbers}) library for high-precision floating point numbers, we have written a new code in an object-oriented programming style. We estimate that the C++ code gives us at least a 20-fold speed-up compared to a Mathematica implementation.

\begin{figure}
    \centering
    \includegraphics[width=0.9\linewidth]{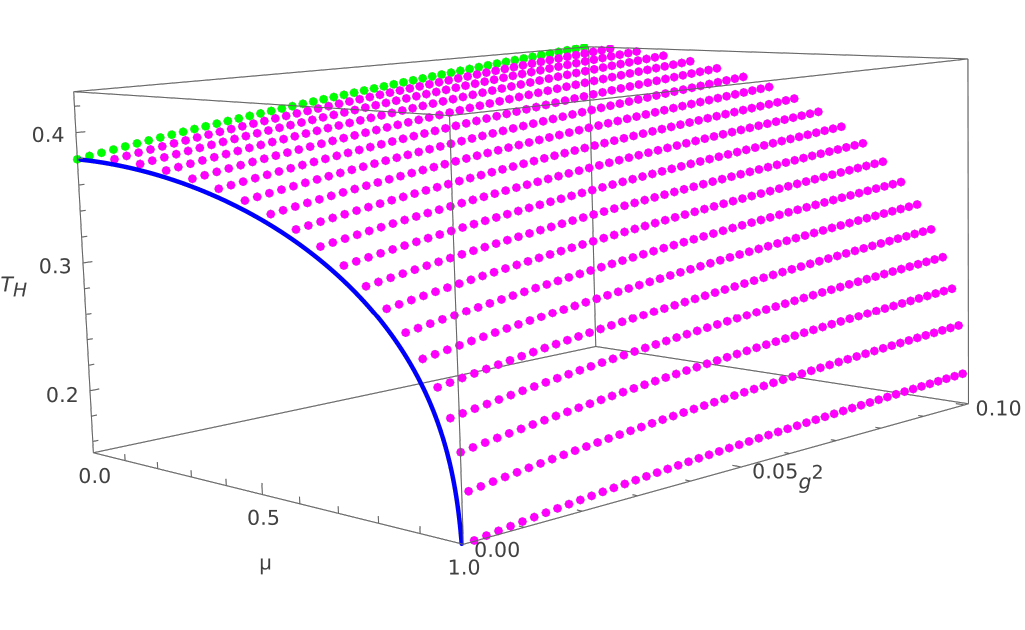}
    \caption{In pink we show the numerical solution of the QSC in the range $g^2\leq0.1$. This is a total of 950 data points. In blue we show again the zero coupling result. In green we show the zero twist results of~\cite{Harmark:2021qma}.}
    \label{fig:weakCoupling}
\end{figure}

We have run the computation for the values $\mu=\sin(\pi\frac{n}{40})$,  $n=1,\dots,19$.  This allows us to compare the numerical results for the Hagedorn temperature to the known weak coupling behavior as well as our analytic conjectures at strong coupling, over a broad range of the chemical potential.    In \autoref{fig:weakCoupling} we show the results of the computation in the range $0\leq g^2\leq0.1$. At zero coupling the numerical data connect well with the zero coupling result from equation \eqref{eq:beads:g0Result} and fit the 1-loop result as can be seen in \autoref{fig:placeholder}. Similarly, the data at zero twist from~\cite{Harmark:2021qma} fit well into the plot, thus giving us an overall smooth picture in the weak coupling regime.

\begin{figure}
    \centering
    \includegraphics[width=1\linewidth]{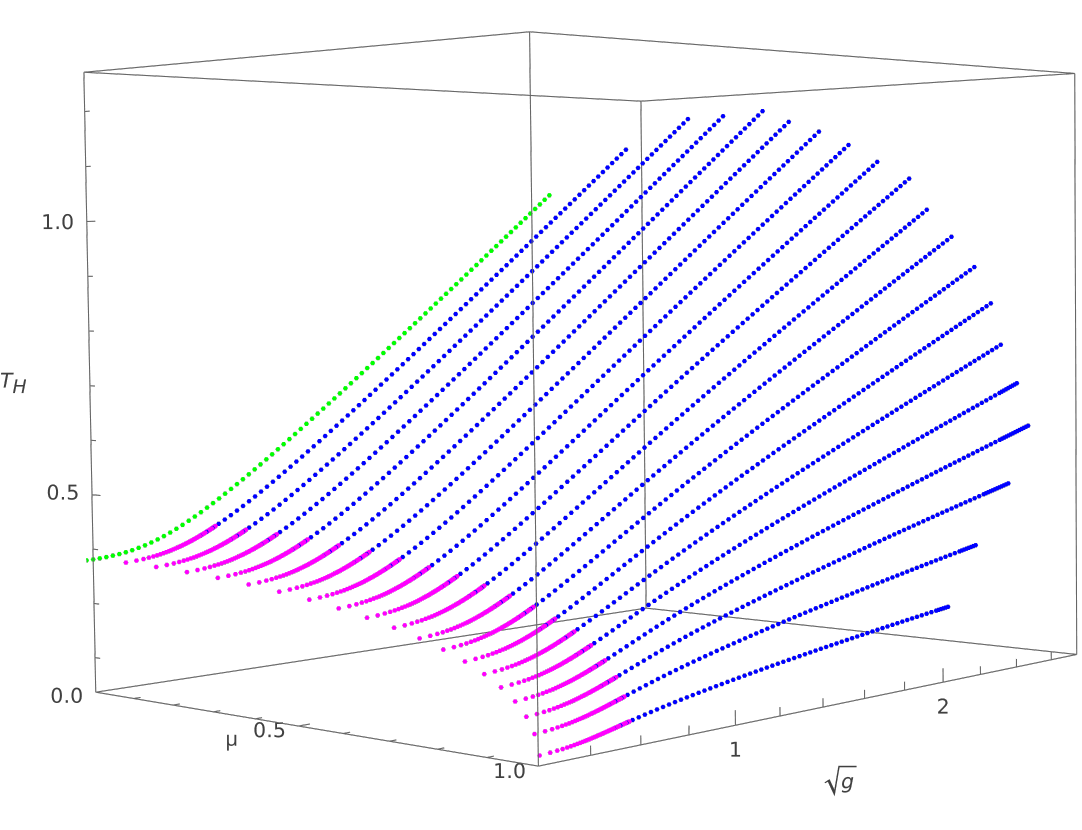}
    \caption{The strong coupling data (1645 points) is shown in blue. In pink we show the weak coupling data from \autoref{fig:weakCoupling} and in green we show the zero twist data from~\cite{Ekhammar:2023glu}.}
    \label{fig:StrongCoupling}
\end{figure}

For most values of the chemical potential we have run our algorithm beyond $\sqrt{g}=2.5$, always reaching at least $\sqrt{g}=2.0$. Figure \ref{fig:StrongCoupling} shows our entire range of numerical results. We estimate the precision of all data to be at least 10 digits. The zero twist data, taken from~\cite{Ekhammar:2023cuj}, fit perfectly into the picture. Going beyond the range of coupling presented here would require a significant amount of computational time. With a cutoff order of 64 for the series expansion of the $\bP$-functions, the duration for a single optimization step running on a single core is already at the order of 1 hour on a desktop (Intel i7 14700 CPU), respectively 3.5 hours on a cluster (2 Intel Xeon Gold 6248R Processors per node). In addition, convergence of the optimization is not always fast, requiring for the most part between 6 and 12 optimization steps. Some help could come from improved extrapolation algorithms. To generate initial data for the optimization we use a polynomial extrapolation from the previous data. This performs poorly for $\mu$ close to 0 respectively 1. We leave improvements to this extrapolation step for future work.

\subsection{Fitting the strong coupling data}

In this subsection, we extrapolate the numerical solution of the quantum spectral curve to get a numerical estimate for the large coupling expansion of the Hagedorn temperature.  We will compare this order by order with the result from the thermal scalar at large twist $\mu\lesssim1$ given in \eqref{THd}. To this end, we will fit expansions in $g^{-1/2}$ to the numerical data and then observe the twist dependence of the fit coefficients. Once we are satisfied that one order of the numerical result matches the strong coupling computation, we subtract the analytic result from the numerical data and we perform a new fit to look at the next subleading order.

\subsubsection{The leading order}

\begin{figure}
    \centering
    \includegraphics[width=0.75\linewidth]{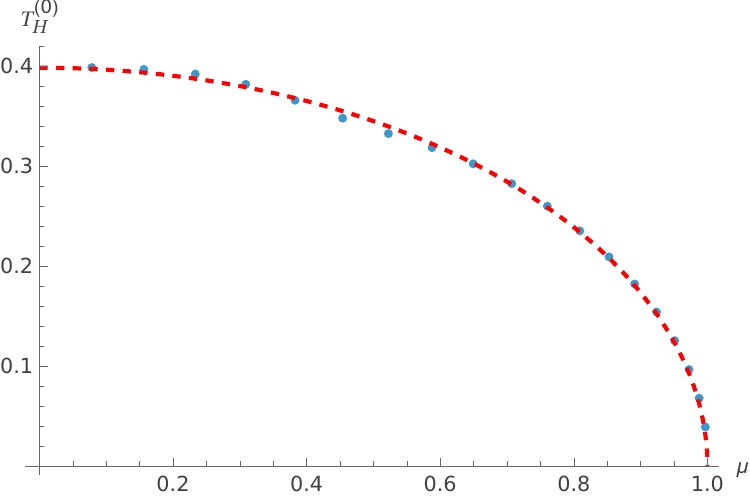}
    \caption{We show the fitted values of the leading order coefficient of the strong coupling fit as a function of the chemical potential. For large $\mu\lesssim1$ it fits very well to the prediction from the thermal scalar (red dashed line).}
    \label{fig:LeadingOrderFit}
\end{figure}

For a fixed value of the chemical potential $\mu$, the leading large $g$ term of the Hagedorn temperature has the form $T_H^{(0)}g^{1/2}$. For the numerical data with $\sqrt{g}\geq1.55$ we fit a function of the form $a_{-1} g^{1/2}+a_0+a_1 g^{-1/2}+a_2g^{-1}$ for each value of the chemical potential $\mu$. We plot the values of $a_{-1}$ as a function of $\mu$ in \autoref{fig:LeadingOrderFit} and compare it against the exact large $g$ result from equation~\eqref{THd}. To quantitatively compare the results we look at $\left(\sqrt{2\pi}\,T_H^{(0)}\right)^2$, which according to equation~\eqref{THd} has the form $\left(\sqrt{2\pi}\,T_H^{(0)}\right)^2=1-\mu^2$. We fit a quartic polynomial in $\mu$ to the square of the numerical results in the range $0.7\leq\mu\leq1$.  This yields \begin{equation}
    \left(\sqrt{2\pi}\,T_H^{(0)}\right)^2\approx 1.006-1.009 \mu ^2-0.007 \mu ^4,
\end{equation} where we have rounded to 3 decimal digits. We see that the numerical result fits  the analytic result within 0.9\,\%. Hence, we are confident that they agree.

\begin{figure}
    \centering
    \includegraphics[width=1\linewidth]{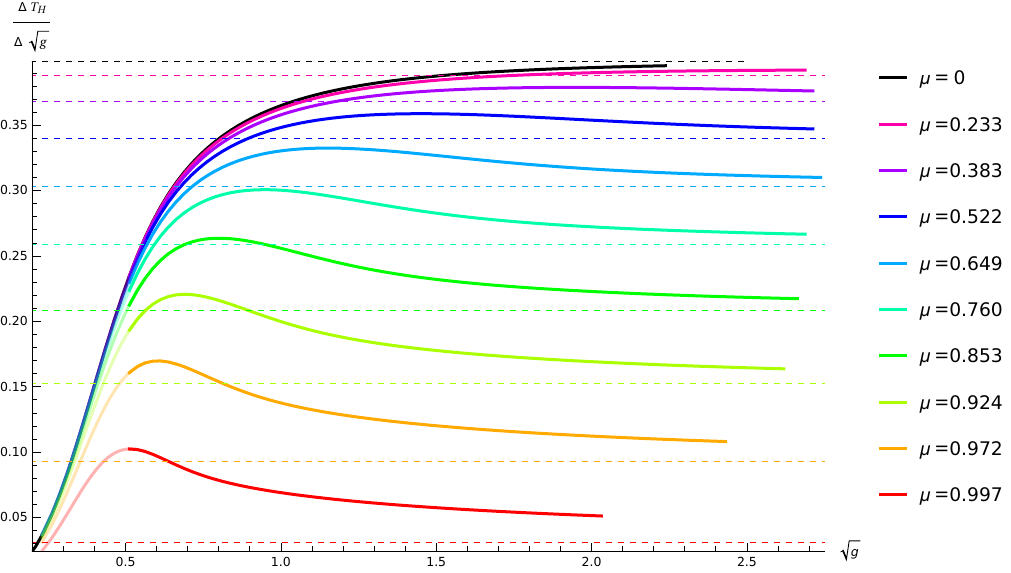}
    \caption{We show the approximation of the first derivative of the Hagedorn temperature as a function of $\sqrt{g}$ for some values of the chemical potential (solid lines). The dashed lines show the value of the leading term in the large $g$ asymptotic expression of the Hagedorn temperature.}
    \label{fig:FirstDerAsymptotics}
\end{figure}

Below $\mu=0.6$ we observe deviations of the numerical results from the large $g$ analytic result for the leading term of the Hagedorn temperature. This can be understood as the first derivative of the Hagedorn temperature $\frac{\partial T_H}{\partial\sqrt{g}}$ is not monotonic. In~\autoref{fig:FirstDerAsymptotics} we show the approximation of this derivative from our numerical data using a finite-difference quotient. For $\mu=0$, this derivative increases monotonically as it asymptotes to $T_H^{(0)}$. For $\mu>0$ the derivative still asymptotes to $T_H^{(0)}$ but we see from the data that it first reaches a maximum before approaching the asymptotic value from above. For $\mu\approx 0.233$ the maximum of the derivative is not reached in the range of couplings we consider. This makes it difficult to determine the correct asymptotic behavior of the Hagedorn temperature from the fitted function. For $\mu\approx0.522$, the derivative of the Hagedorn temperature is decreasing beyond $\sqrt{g}=1.5$ but the inflection point where it starts flattening out again is barely, if at all, visible and thus it is easy to underestimate the asymptotic value.

\subsubsection{The first subleading term}

\begin{figure}
    \centering
    \includegraphics[width=0.75\linewidth]{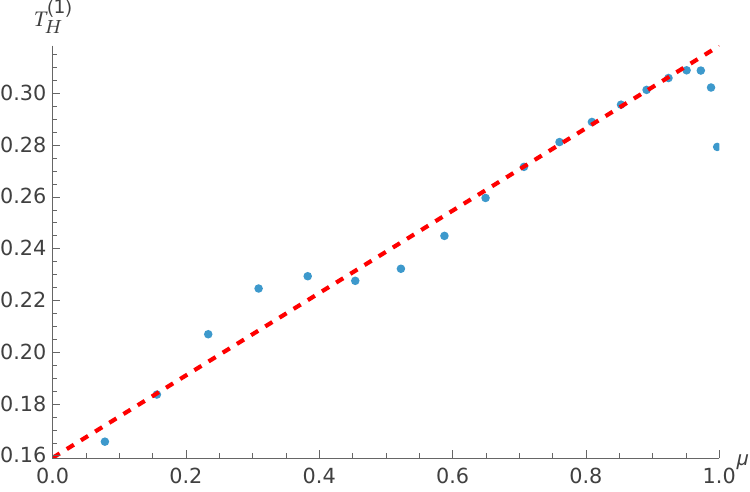}
    \caption{We show the coefficient of the first subleading order of the strong coupling fit as a function of the chemical potential. This fits very well to the prediction from the thermal scalar (red dashed line).}
    \label{fig:NLO_Fit}
\end{figure}

The first subleading term in \eqref{THd} is  $T_H^{(1)}=\frac{1+\mu}{2\pi}$, which is independent of $g$. To compare this with the numerical data, we start by subtracting the leading term in \eqref{THd} from the data. We then fit a function of the form $\sum_{i=0}^3 a_i g^{-i/2}$ to the data for $\sqrt{g}\geq1.55$. The results for the constant term are shown in \autoref{fig:NLO_Fit}. For $0.64<\mu<0.97$ we find that the relative errors between the numeric and analytic result, $\frac{T_H^{(1),n}-T_H^{(1),a}}{T_H^{(1),a}}$ are all below 1.9\,\%, with most being well below 1.0\,\%. Fitting a linear function on this same range of data $0.64<\mu<0.97$, we find \begin{equation}
    T_H^{(1)}\approx \frac{1}{2\pi}(1.017 + 0.973 \mu),
\end{equation} where we have rounded to 3 decimal digits. This agrees  with the analytic result within 2.7\,\%.

\subsubsection{The second subleading order}

\begin{figure}
    \centering
    \includegraphics[width=0.75\linewidth]{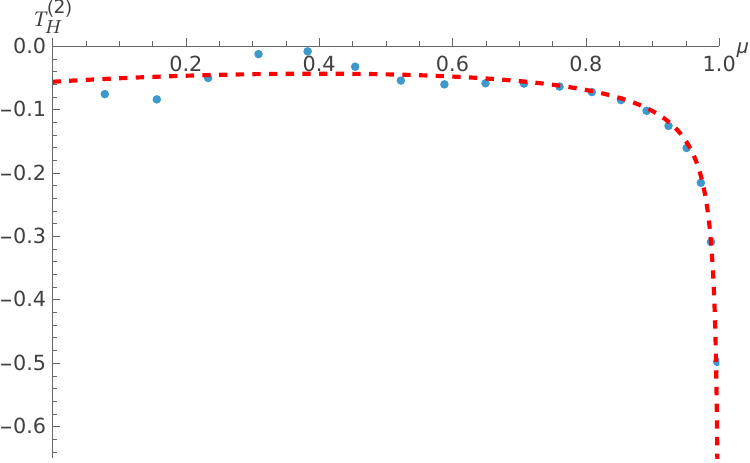}
    \caption{We show the coefficient of the second subleading order of the strong coupling fit as a function of the chemical potential. This fits very well to the prediction from the thermal scalar (red dashed line).}
    \label{fig:N2LO_Fit}
\end{figure}

\begin{figure}
    \centering
    \includegraphics[width=0.75\linewidth]{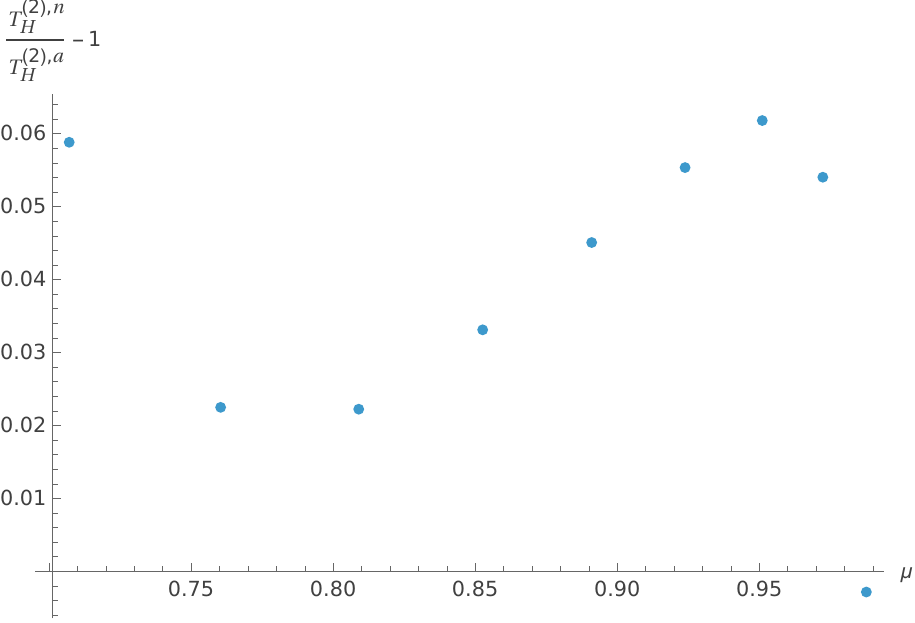}
    \caption{The relative difference of the second subleading term of the Hagedorn temperature}
    \label{fig:N2LO:relDiff}
\end{figure}

The second subleading term in \eqref{THd} is the first order at which the conjecture in \cite{Harmark:2024ioq} plays a role, so the numerical results serve as an important check.
Subtracting the first two orders in \eqref{THd} from the numerical data, we can estimate the second subleading order by fitting a function of the form $\sum_{i=1}^4 a_i g^{-i/2}$ for $\sqrt{g}\geq1.55$. The results are shown in~\autoref{fig:N2LO_Fit}. The analytic result for this term in \eqref{THd} is more involved. Visually, this agrees very well with the numerical data. To quantify this agreement, we compute the relative difference between the numerical and analytical results and plot it in \autoref{fig:N2LO:relDiff}. This is below 6.2\,\% for the range $0.7<\mu<0.98$. In this same range we can also fit a quadratic polynomial in $\mu$ to the rescaled second subleading term, $\sqrt{1-\mu^2}T_H^{(2)}$. 
The result is \begin{align}
   4\sqrt{2\pi^3}\sqrt{1-\mu^2}T_H^{(2)}&\approx-1.758 + 1.861 \mu - 1.705 \mu^2\\
   &=(1+\mu)^2-4(1+\mu^2)\log(2)+0.015 - 0.139 \mu+ 0.068\mu^2,\nonumber
\end{align} where we have rounded to 3 decimal digits. We observe that this fit is very close to the analytic result and gives us confidence that the terms agree.   

\subsubsection{Final remarks on the numerics}

\begin{figure}
    \centering
    \includegraphics[width=0.75\linewidth]{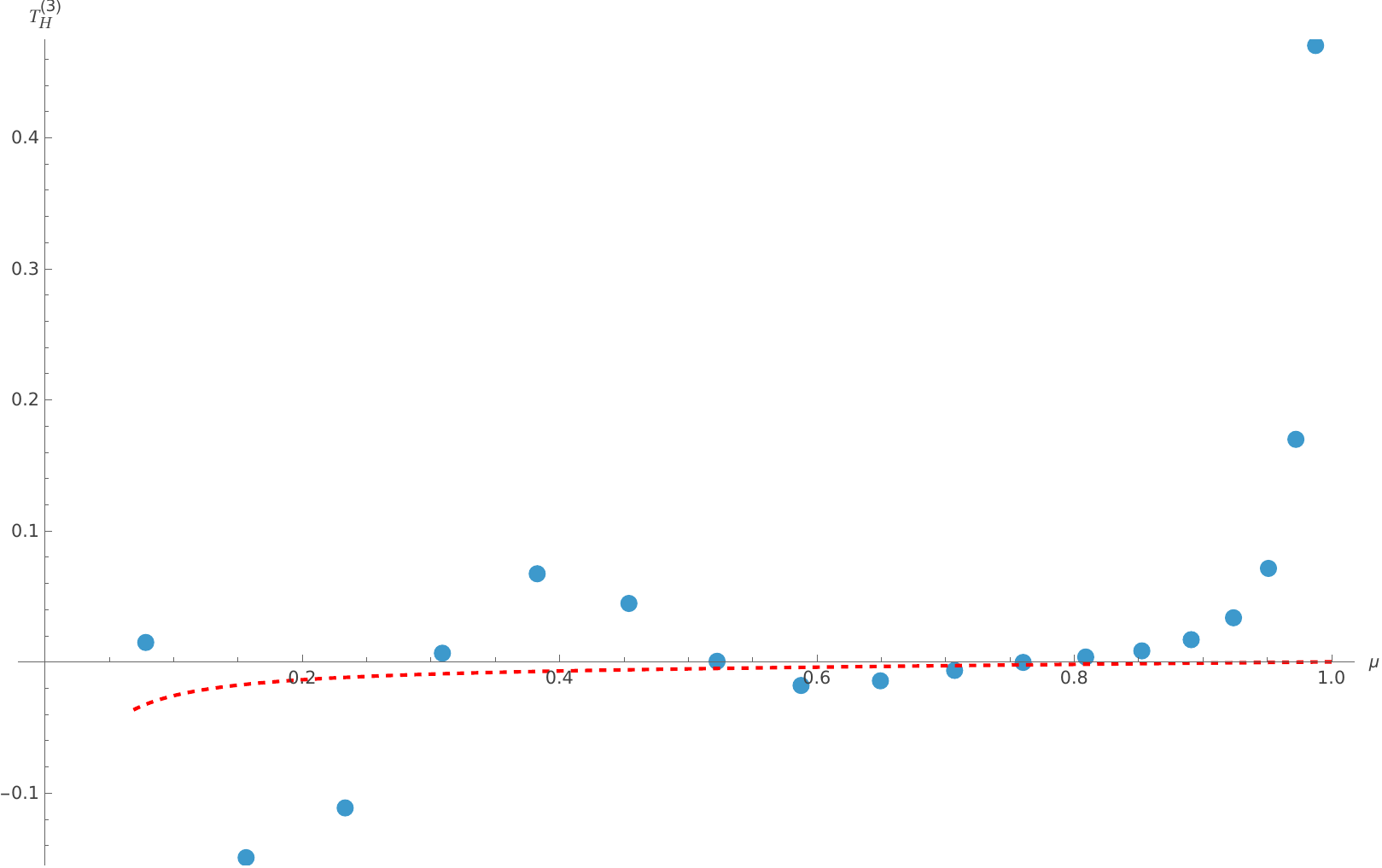}
    \caption{We show the coefficient of the third subleading order of the strong coupling fit as a function of the chemical potential. This doesn't fit well to the prediction from the thermal scalar (red dashed line).}
    \label{fig:N3LO_Fit}
\end{figure}

Unfortunately, our current data are too noisy to study higher order corrections. 
Figure~\ref{fig:N3LO_Fit} shows an attempted fit to the third sub-leading order in \eqref{THd}, which not only shows that the fluctuations are much larger than the predicted value for almost the entire range of $\mu$, but there is also a large tail as $\mu\to1$.  We will need results at much stronger couplings for the entire range of the twists  to confidently fit the data at this order and beyond. A large part of the problem  is that the prediction for $T_H^{(3)}$ in \eqref{THd} is two orders of magnitude smaller than the first three terms over most of the twist range. To compare with figures \ref{fig:LeadingOrderFit}, \ref{fig:NLO_Fit} and \ref{fig:N2LO_Fit}, we observe that  $-0.0055<T_H^{(3)}\leq0$ for $0.5\leq\mu\leq1$.

We have also excluded the $\mu\approx0.997$ results from our comparisons  at strong coupling. As argued at the end of section \ref{muls1}, when  $\mu$ approaches 1 it is appropriate to use the rescaled coupling $\wt{g}=g(1-\mu^2)$. This coupling only reaches $\sqrt{\wt{g}}\approx0.1569$ at $\mu\approx0.997$ in our current data set. Thus,  we have not yet reached the true strong coupling regime for high twist.

\section{Conclusions}\label{sec:Conclusion}

In this paper we have generalized the QSC to find the Hagedorn temperature in the presence of a nonzero chemical potential coupled to one of the $R$-charges.  The QSC results match the previously known weak coupling results, as well as  predictions at strong coupling using a  string wrapped on the thermal circle in $AdS_5$ and localized near a great circle of $S^5$.  The numerical results give further credence to the conjecture in \cite{Harmark:2024ioq} for the structure of the zero-point shift on the string world-sheet.

    There are several further directions that can be explored. One possibility is to use the ideas of~\cite{Ekhammar:2024rfj} to compute the Hagedorn temperature at much higher values of the coupling. This will lead to more precision for the subleading terms of the Hagedorn temperature and thus reveal important information about higher order corrections in string theory.

Another direction is to consider the limiting behavior of the chemical potentials.   In \cite{Gromov:2014bva} it was shown that the QSC can be expanded for small spin around BPS operators, enabling one to find the so-called slope and curvature functions for any value of the coupling. It is tempting to look for a similar simplification in our case by tuning the twist $\mu$. Indeed, we observed that by first twisting all $\bP$-function so that one considers the Witten index and then sending $\mu\rightarrow 1$ the QSC simplifies considerably. It would be interesting to explore this regime further. 

It would also be interesting to use the QSC to numerically compute the Hagedorn temperature at strong coupling for nontrivial chemical potentials coupled to the spins. This will require improved solving techniques, since twisting the exponential asymptotics of the $\bQ_i$\,-functions with the spin-twists leads to numerical instabilities using the current algorithm.

\section*{Acknowledgments}

We thank Aleksey Cherman for pointing out the signifance of \cite{Cherman:2020zea}. C.T. acknowledges funding support from an STFC Consolidated Grant ‘Theoretical Particle Physics at City, University of London’ ST/T000716/1.  J.A.M. thanks the Center for Theoretical Physics -- a Leinweber Institute  at MIT for 
hospitality during the course of this work. The work of S.E. was conducted with funding awarded
by the Swedish Research Council, Vetenskapsrådet, grant VR 2024-00598.

\appendix

\section{The one loop spin twisted Hagedorn temperature}\label{app:OneLoopTwisted}
In this appendix we collect the relevant expressions for the spin twisted Hagedorn temperature. The expectation value of the $\mathcal{N}=4$ dilation operator is
\begin{equation}
\begin{split}
    \expval{D_{2}}& \left(1-y^{2} w z\right)\left(1-y^{2} \frac{w}{z}\right)\left(1-y^{2} \frac{z}{w}\right)\left(1-y^{2} \frac{1}{w z}\right)\\
=&\left[\scalebox{1}{$\frac{w (y+z)^4 \left(4 y z^3 \left(w^2 y^2+y^4-1\right)+z^2 \left(\left(w^4-6\right) y^2-w^2+y^6\right)+6
   y^4 z^4-y^4-4 y^3 z\right)}{\left(w^2-1\right) z^3 \left(z^2-1\right) \left(w z-y^2\right)}\log\left(1-\frac{y^2}{w z}\right)$}\right. \\
    &+\scalebox{1}{$\frac{(y+z)^4 \left(w^6 \left(-y^4\right)-4 w^4 y^3 z+4 y z^3 \left(w^2 y^4+y^2-1\right)+z^2 \left(w^4
   y^6+\left(1-6 w^2\right) y^2-1\right)+6 y^4 z^4\right)}{\left(w^2-1\right) z^3 \left(z^2-1\right)
   \left(w y^2-z\right)}\log\left(1-\frac{w y^2}{z}\right)$}
   \\
   &-\scalebox{1}{$\frac{w (y z+1)^4 \left(y^2 \left(w^4+y^4-y^2\right)-z^2 \left(w^2-6 y^4+6 y^2\right)+4 y^3 z
   \left(w^2+y^2-1\right)-4 y z^3\right)}{\left(w^2-1\right) z^2 \left(z^2-1\right) \left(w-y^2 z\right)}\log\left(1-y^2\frac{w}{z}\right)$}\\
  &\left.+\scalebox{1}{$\frac{(y z+1)^4 \left(y^4 \left(w^6-6 z^2\right)-w^4 y^6+4 \left(w^4-1\right) y^3 z-4 w^2 y^5 z+y^2 \left(6
   w^2 z^2-1\right)+4 y z^3+z^2\right)}{\left(w^2-1\right) z^2 \left(z^2-1\right) \left(w y^2 z-1\right)}\log\left(1-y^2 w z\right)$} \right]\,.
\end{split}
\end{equation} Using this expression the correction to the Hagedorn temperature can then be readily computed through equation \eqref{eq:yH1Loop}.

\section{The algorithm for the numerical solution of the QSC}\label{app:NumAlgorithm}

The numerics algorithm is based on an {\it ansatz} for the $Q$-functions truncated to finite order at large $x$ and $u$ respectively: \begin{align}
	\bP_a&=\begin{pmatrix}
		e^{\frac{\io u \mu}{2T_H}}\bkt{A_1+\sum_{n=1}^{N+1}\frac{c_{1,n} g^n}{(\io x)^n}}\\
		e^{\frac{\io u \mu}{2T_H}}(\io x g)\bkt{A_2+c_{2,0}(\io x g)^{-1}+\sum_{n=1}^{N}\frac{c_{2,n} g^{n-1}}{(\io x)^{n+1}}}\\
		e^{-\frac{\io u \mu}{2T_H}}\bkt{A_3+\sum_{n=1}^{N+1}\frac{c_{3,n} g^n}{(\io x)^n}}\\
		e^{-\frac{\io u \mu}{2T_H}}(\io xg)\bkt{ A_4+c_{4,0}(\io xg)^{-1}+\sum_{n=1}^{N}\frac{c_{4,n} g^{n-1}}{(\io x)^{n+1}}}
	\end{pmatrix} =y^{\io u \mu t_a}\sum_{n=-s_a}^{N+1-s_a} c_{a,n}\bkt{\frac{g}{\io x}}^n
    \end{align}
    \begin{align}
	Q_{a|i}&=\io y^{\io u (\mu t_a+\wt{t}_i)}e^{\pi u \wt{t}_i} \sum_{n=0}^N b_{a|i,n}(\io u)^{s_a+\wt{s}_i-n}, & \bQ_i&
    =e^{\pi u \wt{t}_i}y^{\io u \wt{t}_i} (iu)^{\wt{s}_i}(B_i+\mathcal{O}(u^{-1})).
\end{align} Here we use $t_a=\wt{t}_i=(1,1,-1,-1)$, $s_a=\wt{s}_i=(0,1,0,1)$ and the Zhukovsky variable $x=\frac{1}{2g}(u+\sqrt{u^2-4g^2})$. To fix the gauge we choose equations \eqref{eq:gaugeChoice} and \eqref{eq:QaiLeadCoeffs} to hold for all values of coupling, i.e. with $y_0$ replaced by $y=\exp(\frac{1}{2T_H})$.

The numerical problem is much simplified by symmetry relations for the $\bP$- and $\bQ$-functions: \begin{align}
    \bP_a(-x)&=(-1)^{a-1}\bP_{a+2}(x), & \bQ_i^\uparrow(-u)=(-1)^{i-1}\bQ_{i+2}^\downarrow(u),
\end{align} for $a=1,2$ respectively $i=1,2$, and similar relations for $a,i=3,4$. For the $\bQ$-functions this symmetry relates the lower half-plane analytic function $\bQ_i^\uparrow$ to the upper half-plane analytic function $\bQ_i^\downarrow$. For the coefficients of the $\bP$-functions this symmetry translates into the relations \begin{align}
	c_{4,n}&=(-1)^{n+1}c_{2,n}, & c_{3,n}&=(-1)^{n}c_{1,n},
\end{align} and for the coefficients of the $Q_{a|i}$ we get the relations \begin{align}
	b_{4|1,n}&=(-1)^{n +1}b_{2|3,n}, & b_{4|2,n}&=(-1)^{n +1}b_{2|4,n},\nonumber\\
	b_{4|3,n}&=(-1)^{n +1}b_{2|1,n}, & b_{4|4,n}&=(-1)^{n +1}b_{2|2,n},\nonumber\\
	b_{3|1,n}&=(-1)^{n +1}b_{1|3,n}, & b_{3|2,n}&=(-1)^{n +1}b_{1|4,n},\label{eq:app:symQai}\\
	b_{3|3,n}&=(-1)^{n +1}b_{1|1,n}, & b_{3|4,n}&=(-1)^{n +1}b_{1|2,n}.\nonumber
\end{align} This symmetry thus halves the number of coefficients that need to be fixed.

With the {\it ansatz} fixed we look at the $QQ$-relation \begin{equation}
    Q_{a|i}^+-Q_{a|i}^-=-\bP_a \bP^bQ_{b|i}^+\,.
\end{equation} This can be seen as a linear system of equations for the coefficients $b_{a|i,n}$ of the $Q_{a|i}$ in terms of the coefficients $c_{a|n}$ of the $\bP_a$. In more detail, these relations can be expanded to take the form
\begin{align}
	\begin{split}
		&\sum_{m=-\infty}^{s_a+\wt{s}_i}(\io u)^m\sum_{l=0}^{-m+s_a+\wt{s}_i}\begin{pmatrix}
		-l+s_a+\wt{s}_i\\ -l-m+s_a+\wt{s}_i
	\end{pmatrix}\\
	&\hspace{120pt}\times\bkt{\bkt{-1}^{-l-m+s_a+\wt{s}_i}-(-1)^{\wt{t}_i}y^{(t_a\mu+\wt{t}_i)}}2^{l+m-s_a-\wt{s}_i}b_{a|i,l}
	\end{split}\label{eq:QQexpanded}\\
	 \begin{split}
	 	&=-y^{\frac{1}{2}\mu(t_a-t_b)}\sum_{m=-\infty}^{s_a+s^b}\frac{\sum_{k=-s_a}^{-m+s^b}c_{a,k}c\indices{^b_{,-m-k}}}{g^{2m}}\sum_{n=-\infty}^{s_b+\wt{s}_i}(\io u)^{n+m}\\
	&\hspace{120pt}\times\sum_{p=0}^{-n+s_b+\wt{s}_i}\begin{pmatrix} -p+s_b+\wt{s}_i\\-p-n+s_b+\wt{s}_i\end{pmatrix}(-2)^{n+p-s_b-\wt{s}_i}b_{b|i,p}\sum_{l=0}^\infty\frac{(4g^2)^l}{(iu)^{2l}}d_{m,l},
	\end{split}\nonumber
\end{align} with
\begin{align*}
    \ d_{n,l}=\sum_{m=0}^l \frac{1}{2^{m}}\begin{pmatrix} n\\ m\end{pmatrix}\sum_{k=0}^m (-1)^{m-k} \begin{pmatrix} m\\ k\end{pmatrix}\begin{pmatrix} \frac{k}{2}\\ l\end{pmatrix}.
\end{align*} In table~\ref{tab:NumAlg:orders} we show for given values of $a,i$ at which orders  in $u$ we solve the $QQ$-relation~\eqref{eq:QQexpanded} and count the number of equations that this provides. From the order $u^2$ term of the constraint $Q_{a|i}Q^{b|i}=-\delta_a^b$, we also get the equation \begin{equation}\label{eq:inverseConstraint}
	-b_{2|2,0} b_{2|1,1}+b_{2|1,0} b_{2|2,1}-b_{2|4,0} b_{2|3,1}+b_{2|3,0} b_{2|4,1}=0,
\end{equation} 
which is linear in the subleading terms $b_{a|i,1}$ of $Q_{a|i}$. In total we have $8N$ equations and thus we can solve this linear system to compute the first $N$ subleading orders of the large $u$ expansion of the $Q_{a|i}$. Given a set of values for $c_{a,n}$ and $y$, this solution can be found numerically. The $QQ$-relation also fixes $c_{2,0}$ and $c_{2,1}$ in terms of the other $c_{a,n}$ and $y$. Thus we have in total $(2(N+1)+1)-2=2N+1$ parameters that need to be fixed.

\begin{table}
    \centering
    \begin{tabular}{c|c|c|c|c}
		a & i & highest order & lowest order & \# of equations\\
		\hline
		2 & 2,4 & 2 & -N+3 & 2N\\
		  & 1,3 & 1 & -N+2 & 2N\\
		\hline
		4 & 2 & 1 & -N+2 & N\\
		  & 1 & 0 & -N+1 & N\\
		\hline
		1 & 2 & -4 & -N-2 & N-1\\
		  & 1 & -4 & -N-3 & N 
	\end{tabular}
    \caption{The orders of $u$ for which we solve the $QQ$-relation~\eqref{eq:QQexpanded} for various values of $a,i$. These provide $8N-1$ linear equations for the coefficients $b_{a|i,n}$.}
    \label{tab:NumAlg:orders}
\end{table}

We choose the set of points $I=\{-2g\cos(\frac{2n+1}{2K}\pi)\}_{0\leq n\leq K-1}$ for $K=2N+2$. For a given set of values of $c_{a,n}$ and $y$ we wish to test the gluing condition \eqref{eq:gluing} at these points. To this end we start by evaluating $Q_{a|i}$ at the points $\wt{u}+\io(U+\frac{1}{2})$ for $\wt{u}\in I$ and $U$ a large integer using its large $u$ expansion. Then we can iterate the $QQ$-relation in the form \begin{equation}
    Q_{a|i}^-=(\delta_a^b+\bP_a\bP^b)Q_{a|i}^+\,,
\end{equation} to get the values of $Q_{a|i}^+$ at all the points in $I$. Finally, being careful in evaluating $\bP_a(u+\io 0)$ on the branch cut, we can use $\bQ_i=-\bP^a Q_{a|i}^+$ to get the values $\bQ_i^\downarrow(\wt{u}+\io 0)$ for $\wt{u}\in I$. For an exact solution of the quantum spectral curve, the function
\begin{equation}
    F(y,\{c_{a,n}\})=\sum_{\wt{u}\in I}\sum_{i=1}^4\left\vert \frac{\bQ_i^\downarrow(\wt{u}+\io 0)}{(-1)^{\wt{s}_i}e^{\wt{t}_i\pi u}\bQ_i^\downarrow(-\wt{u}+\io 0)}-1\right\vert^2
\end{equation} 
is identically zero due to the gluing condition \eqref{eq:gluing}.

The aim of the numerical algorithm is to find values of $y$ and $\{c_{a,n}\}$ such that $F$ is approximately zero. To find this approximate zero we use a damped Newton algorithm. At low values of the coupling we initialize the algorithm with random values for $\{c_{a,n}\}$ and the zero coupling value of $y$. For stronger coupling we use extrapolation to generate initial data for the optimization algorithm.

\vfill\eject
\bibliographystyle{JHEP}
\bibliography{refs}

\end{document}